\documentstyle[psfig,aps,pre]{revtex}

\newcommand {\be}{\begin{equation}}
\newcommand {\ee}{\end{equation}}
\newcommand {\bea}{\begin{eqnarray}}
\newcommand {\eea}{\end{eqnarray}}
\newcommand{\va}{\vec{a}}
\newcommand{\vf}{\vec{f}}
\newcommand{\vx}{\vec{x}}
\newcommand{\vv}{\vec{v}}
\newcommand{\tvv}{\tilde{\vec{v}}}
\newcommand{\tx}{\tilde{x}}

\begin{document}
\title{Reconstruction of systems with delayed feedback: (II) Application}
\author{Martin J.\ B\"unner$^1$, Marco Ciofini$^1$, Antonio Giaquinta$^1$, 
Rainer Hegger$^2$}
\author{Holger Kantz$^2$, Riccardo Meucci$^1$, and Antonio Politi$^{1,3}$}
\address{(1) Istituto Nazionale di Ottica Applicata\\ Largo E.\ Fermi 6,
50125 Firenze, Italy}
\address{(2) Max--Planck--Institut f\"ur Physik komplexer Systeme\\
N\"othnitzer Str.\ 38, 01187 Dresden, Germany}
\address{(3) INFM - Unit\`a di Firenze}

\date{\today}
\maketitle

\begin{abstract}
We apply a recently proposed method for the analysis of time series
from systems with delayed feedback to experimental data generated by a 
$CO_2$ laser. The method is able to estimate the delay time 
with an error of the order of the sampling interval, while an approach
based on the peaks of either the autocorrelation function, or the time 
delayed mutual information would yield systematically larger values.
We reconstruct rather accurately the equations of motion and, in turn, estimate 
the Lyapunov spectrum even for rather high dimensional attractors. 
By comparing models constructed for different ``embedding dimensions'' with
the original data, we are able to find the minimal faitfhful model.
For short delays, the results of our procedure have been cross-checked 
using a conventional Takens time-delay embedding. For large 
delays, the standard analysis is inapplicable since the dynamics becomes 
hyperchaotic. In such a regime we provide the first experimental evidence
that the Lyapunov spectrum, rescaled according to the delay time, is 
independent of the delay time itself. This is in full analogy with the
independence of the system size found in spatially extended systems. 
\end{abstract}

\section{Introduction}
\label{sec.intro}
\begin{figure}
\centerline{\psfig{file=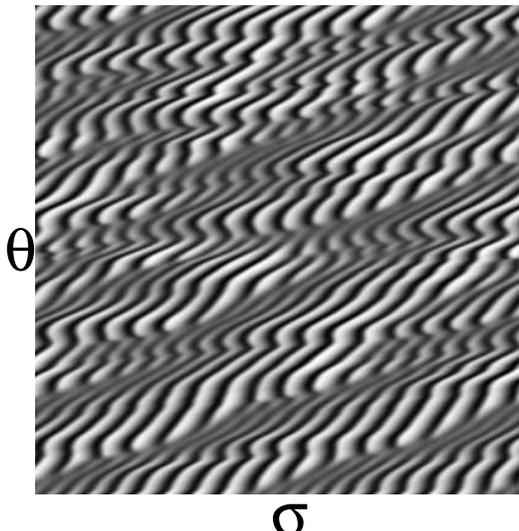,width=7cm,angle=270}}
\caption{Space-time representation of the $CO_2$ laser intensity
for a delay $\tau_0 = 400 \mu s$ and for $\tau_1>\tau_0$ chosen so
as to optimize the coherence of the vertical structures.}
\label{fig.r400-st}
\end{figure}

In many physical, biological, chemical and technical systems, feedback
loops involve a time delay. Typical examples include population
dynamics, where individuals participate in the reproduction of a
species only after maturation, or spatially extended systems where
signals have to cover distances with finite velocities (e.g.\
reflections in optical fibre networks coupled back to some light
source). Within this rather broad class of systems, one can find the 
Mackey-Glass equation \cite{Mackey77} modelling the production of red 
blood cells and the Lang-Kobayashi equations \cite{Lang} describing 
semiconductor lasers with optical feedback. To be more specific, let us
consider the following class of delayed differential equations (DDE) 
\be
\dot{\vx}(t)=\vf\left(\vx(t),x_l(t-\tau_0)\right)\;,
\label{dde-def.eq}
\ee
where $\vx\in{\mathbf R}^d$ and $x_l(t-\tau_0)$ is a single component fed
back into the system with a fixed delay $\tau_0$. Although
Eq.~(\ref{dde-def.eq}) appears to be very simple, the phase space of the 
system is infinite dimensional, namely the direct product 
${\mathbf R}^d\otimes
C_1([-\tau_0,0[,{\mathbf R})$, where $C_1([-\tau_0,0[,{\mathbf R})$ 
is the space of differentiable functions from the interval 
$[-\tau_0,0[$ to ${\mathbf R}$. Accordingly, such systems are in principle
able to sustain arbitrarily high-dimensional dynamics.
One way to elucidate this point is by interpreting delayed dynamical 
systems as spatially extended ones with the introduction of
a two-variable representation of the time
\be
\hat{\vx}(\sigma,\theta)= \vx(t=\sigma + \theta \tau_1),
\label{eq.space-time}
\ee
where $\sigma$ is a continuous space-like variable bounded between 0 and
$\tau_1 \approx \tau_0$, while $\theta$ is a discrete temporal variable 
numbering the delay units \cite{AGLM92,GiPo96}. In fact, this
representation allows interpreting the long range interaction due to the 
delay as a short range interaction along the $\theta$ direction, since 
$y(t-\tau_1) \equiv y(\sigma, \theta -1)$. The meaningfulness of this
representation can be appreciated in Fig. \ref{fig.r400-st}, where 
the propagation of coherent structures across different delay units is
clearly visible in data taken from the laser experiment described in this 
paper (see the next section for a description of the system).

The most direct approach to the investigation of the dynamics of an 
experimental system consists in analysing the time record of a suitable 
observable. In the first part of this work \cite{Buenner99}, we discussed 
in detail the theoretical framework for the analysis of time-delayed feedback 
systems. In fact, it was necessary to develop a specific theory to 
reconstruct the equations of motion, since the high complexity of 
typical time-delayed feedback systems usually prevents the implementation
of the standard embedding techniques\cite{Takens80,Sauer91a}. 

We are able to identify the deterministic structure underlying one such time 
series by adopting the ideas developed by Casdagli~\cite{Casdagli92} 
for input--output systems. These are systems such as
\be\label{i_o_def.eq}
\vx_{n+1}=\vf\left(\vx_n,\epsilon_n\right)\;,
\ee
where $\vx_n\in{\mathbf R}^d$ is the state vector and $\epsilon_n$ is an 
additional time-dependent input. If a time series $\{y_n\}$ of some scalar 
observable $y(\vx)$ is available together with the (simultaneously recorded)
input $\{\epsilon_n\}$, then Casdagli argued and more recently Stark et
al.~\cite{SBDH97} have proven that the use of vectors of the form
\be\label{vv.eq}
\vv_n=(y_n,y_{n-1},\ldots,y_{n-m+1},\epsilon_n,\epsilon_{n-1},\ldots,
\epsilon_{n-m+1})
\ee
provide a proper embedding for $m>2d$. Such vectors unambiguously define
the state of the system and in principle their knowledge uniquely determines 
the next observation $y_{n+1}$.

A discrete-time version of Eq.~(\ref{dde-def.eq}) has formally the same
structure as an input-output system with the only difference that the input 
$\epsilon_n$ has to be replaced by the time-delayed value of the 
variable $x_l$ (see also \cite{Hegger98,Buenner97}). 
Therefore, a time series of $x_l$ allows to form vectors which are equivalent 
to those in Eq.~(\ref{vv.eq}). The same is possible also when the recorded
variable does not coincide with the feedback one, although a higher-dimensional
state space is required in this case (see Ref.~\cite{Buenner99}).
More precisely, given the signal $x_l(t)$, we introduce the vectors
\be\label{eq:ourspace}
\vv_n(m,\tau)=(y_n,y_{n-1},\ldots,y_{n-m+1},
y_{n-T},\ldots,y_{n-T-m+1}),
\ee
where $y_n=x_l(t=n\delta t)$ ($\delta t$ is the sampling time) and 
the integer $T$ corresponds to the physical time $\tau=T\delta t$. 
In Ref.~\cite{Buenner99}, it has been shown that for $m>2d$ and 
$T=\tau_0/\delta t$, the knowledge of these vectors is (almost) sufficient
to determine the future dynamics. In fact, some approximations arise because 
of the infinite-dimensional phase-space of the original continuous-time 
system that is replaced by a finite-dimensional space in the 
discrete-time representation.
However, we have already seen in numerical simulations \cite{Buenner99} 
and we shall confirm here by analysing experimental data that such
approximations are harmless if the sampling time is sufficiently short.

A conceptually more satisfying reconstruction could be obtained by
preserving the continuity of time, but this would be definitely less practical
since the computation of derivatives from time-series data is a numerically
unstable procedure that drastically increases any kind of noise.

The simplest convincing evidence that a given vector $\vv_n(m,\tau)$
(for the rest of this work we will assume $\delta t=1$ and thus
identify $T$ and $\tau$) provides
a faitfhful reconstruction of the dynamics is through a good
forecast of the next observation $y_{n+1}$. In practice, we introduce the
following ansatz for the dynamics in the state space
\be
y_{n+1}=h_p\left(\vv_n(m,\tau)\right)\;,
\label{def-ansatz.eq}
\ee
where the function $h_p$ belongs to some class of parametrized functions.
The unknown parameters $p$ are determined by minimizing 
the prediction error 
\be
\sigma_m =\sum_n (y_{n+1}-h_p\left(\vv_n(m,\tau)\right))^2
\label{pred-err}
\ee
with respect to the parameters $p$. A reasonable class of functions
from the point of view of robustness, flexibility and numerical ease
are local linear functions, 
\be
y_{n+1}=b_n+\va_n\vv_n(m,\tau)\;,
\ee
where the parameters $\va_n$ and $b_n$ are determined for each individual 
$\vv_n$ by fitting the behaviour in a suitable neighbourhood of $\vv_n$ 
\cite{FarmerSidorowich}.
The resulting average forecast error (FCE) is thus computed as a function of
$\tau$ for different embedding dimensions $m$. Its minima are used to 
determine the optimal choice of $m$ and $\tau$. 

The representation of the dynamics by local linear maps allows most
easily to compute the Lyapunov spectrum of the system, which is then
approximated by $\tau+m$ exponents. It has been verified in many numerical
simulations and there are good arguments showing that these are
(approximations of) the largest $m+\tau$ exponents of the original system
\cite{Hegger99}. The implementation of this procedure will become definitely 
more transparent in Sec. III and IV where we discuss its application.

This paper is a case study on experimental data from a laser system to 
illustrate how far the investigation of a rather high dimensional dynamics 
can go when suitable methods are employed. Difficulties and limitations
will be illustrated, but much more the overall power of the
method will be proved.

In Sec. II, we describe the experimental set-up. In Secs. III and IV we 
study the low- and high-dimensional dynamics exhibited by
the experiment for different delay times, while in Sec.~V we present
methods for the validation of the results. We will be able to obtain
equations of motion that will be used for short time predictions, for the
computation of the Lyapunov spectrum and for the generation of new
time series, whose properties (such as the invariant measure) will be
compared with the experimental data. Finally, we vary the
time lag $\tau$ in the model equations obtained for a given delay time. 
This is to test the stability of the model that should be independent of
the delay (as long as the equations of motion do not depend explicitely 
on $\tau$). The comparision of such numerically generated data with the 
experimental data obtained for the corresponding delay further confirm the 
validity of the reconstructed model. In summary, the following sections 
will demonstrate that the method proposed in \cite{Buenner99} can be 
successfully employed in real experiments.

\section{The experimental setup}
\label{sec.exp-setup}
The experimental setup employs a single-mode CO$_2$ laser
with an electro-optic feedback on the cavity losses (Fig.~2).
The feedback consists in a signal proportional to the laser output intensity
which, after a suitable delay and amplification, drives an intracavity
electro-optic modulator. The intensity is detected with a fast
Hg-Cd-Te photodiode and the delay line is realized using fast
(2 MHz) and accurate (12 bits) A/D (analog-to-digital) and D/A
converters allowing variations of the delay $\tau_0$ up to 130 ms
with 0.5 $\mu$s resolution. The high-voltage amplifier adds to the
delayed signal a continuous voltage level B which, once $\tau_0$ is
fixed, acts as a control parameter. The overall error in the delay time 
is dominated by the delay line uncertainty, $\delta \tau_0 = \pm 0.5 \mu$s.

\begin{figure}
\centerline{\psfig{file=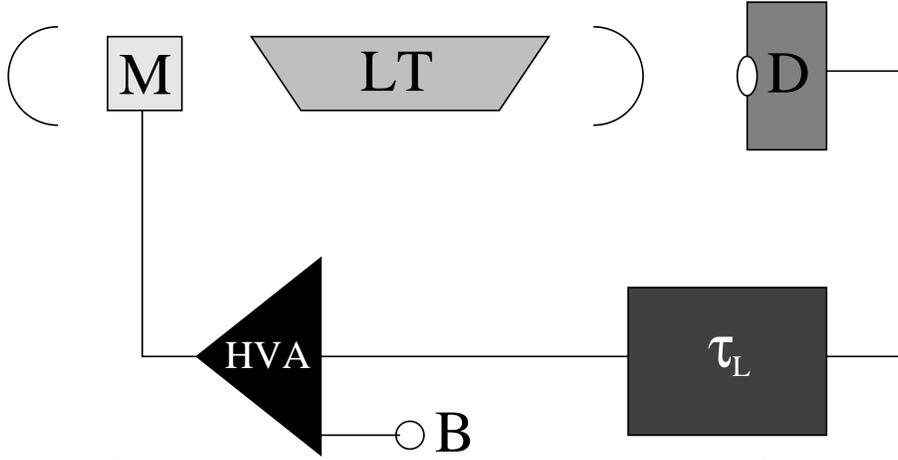,width=12cm,angle=270}}
\caption{Scheme of the experimental setup: M, electro-optic
modulator; LT, laser tube; D, detector; $\tau_L$, variable delay
line; HVA, high voltage amplifier (amplification factor = 10000); 
B, bias voltage input.}
\label{fig.exp_setup}
\end{figure}

The introduction of a delay $\tau_0$ in the feedback loop induces,
upon increasing the bias B, a Hopf bifurcation with the
frequency $f \approx 40.0$ kHz and the
appearance of other incommensurate frequencies leading to
the chaotic regime.
Anyway, if $\tau_0$ is of the order of the characteristic
period ($\sim$ 25 $\mu$s), the fractal
dimension of the chaotic attractor remains between 2 and 3 
\cite{Farmer82,Dorizzi87,Giacomelli91}. 
Since the aim of the present work is to study the transition
to a high dimensional chaotic regime characterized by more than
one positive Lyapunov exponent, we have explored a delay
range between $\tau_0=50 \mu$s and $\tau_0=400 \mu$s.
For these $\tau_0$ values, the system shows the same qualitative 
behaviour. Superimposed to the Hopf oscillation, we observe a deep 
modulation paced by the inverse of the
delay $\tau_0$, while the attractor becomes chaotic. A typical time
sequence is shown in Fig. 2 (for B=250 V) together with the 
corresponding broad-band power spectrum. A further increase of B leads to
a collapse of the attractor into a stable limit cycle. 

\begin{figure}
\centerline{\psfig{file=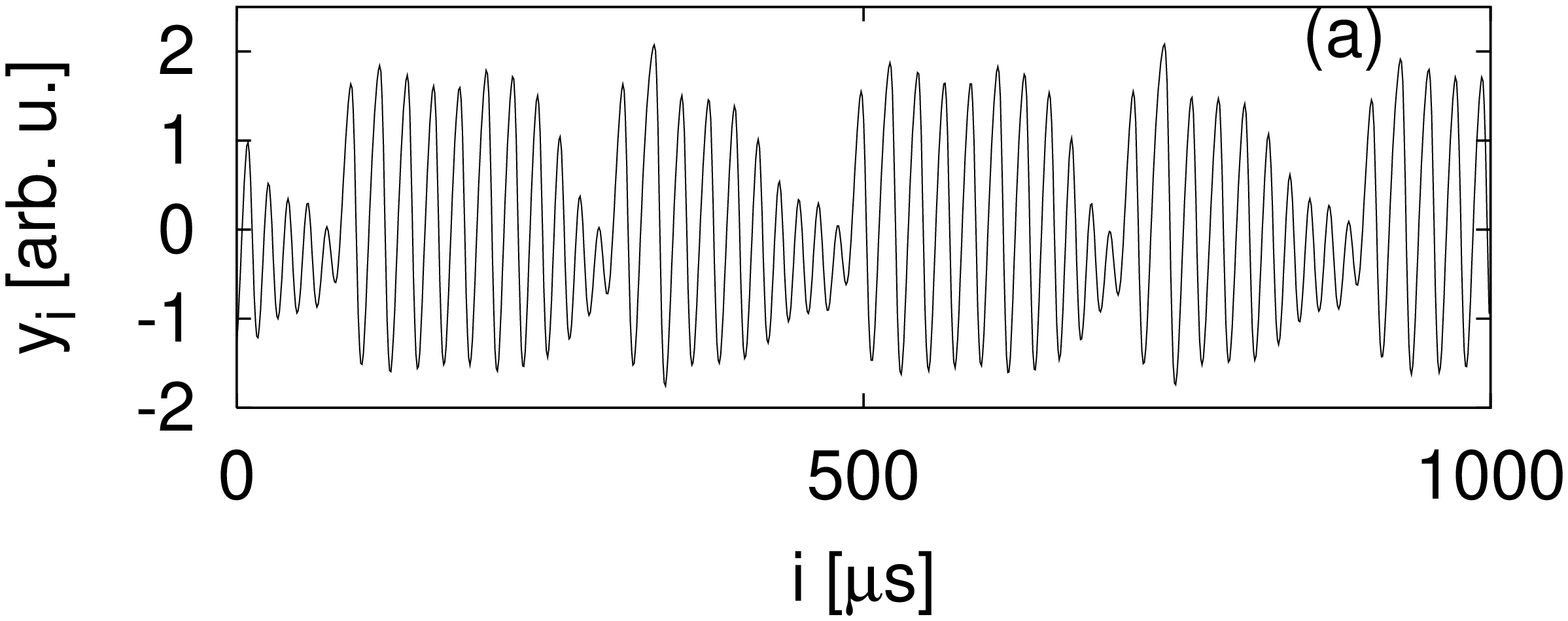,width=12cm,angle=270}}
\centerline{\psfig{file=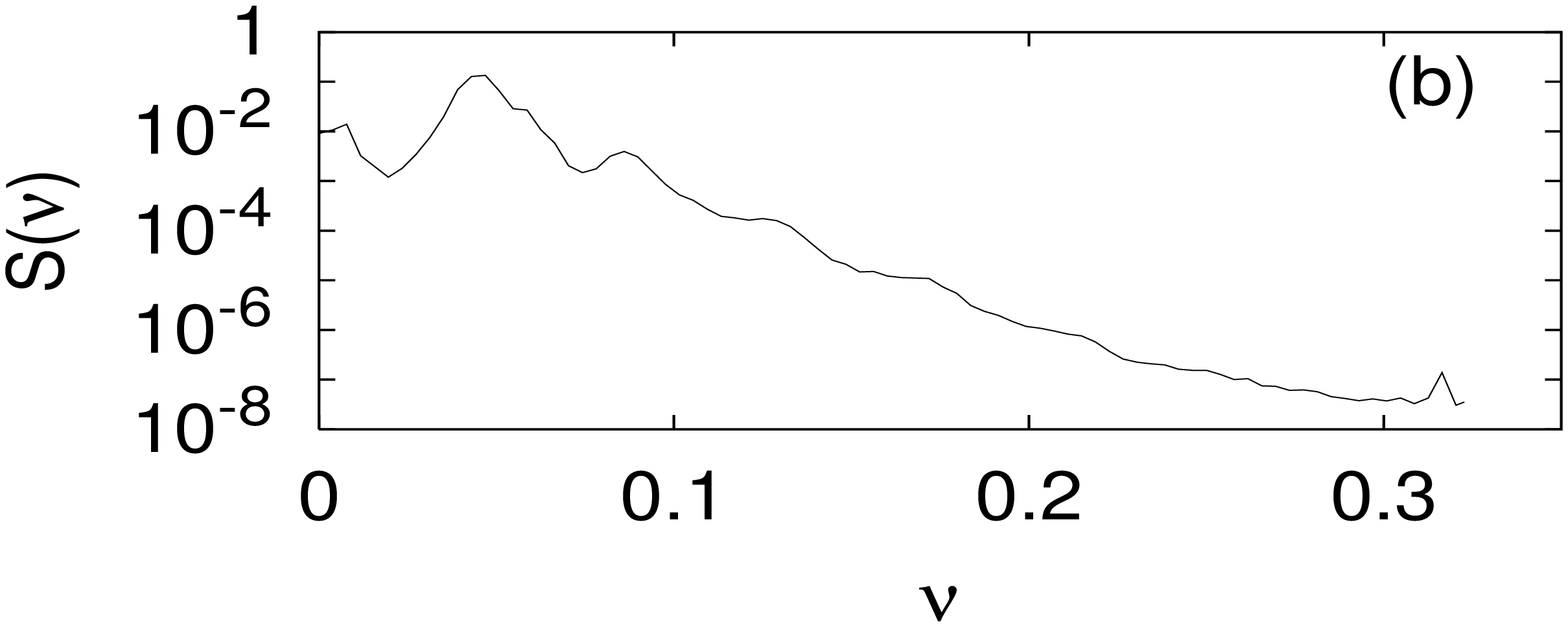,width=12cm,angle=270}}
\caption{(a) High dimensional time series of the $CO_2$-laser in the
long delay regime, $\tau_0=400 \mu$s. (b) Power spectrum of such a
time series.}   
\label{fig.r400-ts}
\end{figure}

The chaotic time series analyzed in the next sections have been
acquired by using a 12 bits A/D converter (LeCroy 6810) with sampling
rate 1.0 $\mu$s and one million data memory. 

The laser dynamics is influenced by ``intrinsic'' noise (dynamical
noise) of the overall experimental setup.  The latter is dominated by
the noise of the electronic part of the feedback loop, which amount to
approximately $3$ bits, i.e. about half a percent. This noise
sets a lower bound for the prediction error computed later.

The simplest approach to model the dynamics of a single mode
homogeneously broadened $CO_2$-laser is based on two rate equations
for the laser intensity and the population inversion between the two
relevant levels. However, this simple model is not adequate to fit data coming
from experiments in the Q-switching regime, or in chaotic regimes
obtained by sinusoidal modulation of the cavity losses, or 
with electro-optic feedback such as investigated in this paper.

A four-level molecular scheme, taking effectively into account the coupling 
between the resonant and some rotational levels provides a more accurate
description of the laser dynamics. Nevertheless, the resulting model
(a six-component DDE \cite{Varone95}) cannot be easily used to reproduce
the behaviour of a $CO_2$-laser with electro-optical feedback because of
the necessity of a fine tuning of the various parameters.

From the huge set of different measurements, we have selected  three
data sets we want to concentrate on in the rest of this paper. They
mainly differ in the delay time, although other parameters of the
laser were slightly modified too in order to keep the system in the 
same dynamical regime. The corresponding delays are 
$\tau_0=(50 \pm 1) \mu s$,
$\tau_0=(150 \pm 1) \mu s$ and $\tau_0=(400 \pm 1) \mu s$.
The signal is sampled at times $t_i = i \delta t$ with the sampling time 
$\delta t =1.0 \mu s$ in all three measurements. The laser intensity is 
measured with a 12-bit resolution (0-4095) and afterwards
normalized such that $\langle y\rangle=0$, $\langle y^2\rangle=1$ in
each individual time series. Each of the three data sets consists of
$10^6$ points.

\section{The low dimensional case}
\label{sec.r50}
We start the analysis of the laser system from the data set
$\{y_i\}$ created with a sufficiently short delay time ($\tau_0=50$ $\mu$s)
to ensure a low-dimensional dynamics. This has the advantage that we can 
cross-check our results with the outcome of standard embedding techniques.

\begin{figure}[ht]
\centerline{\psfig{file=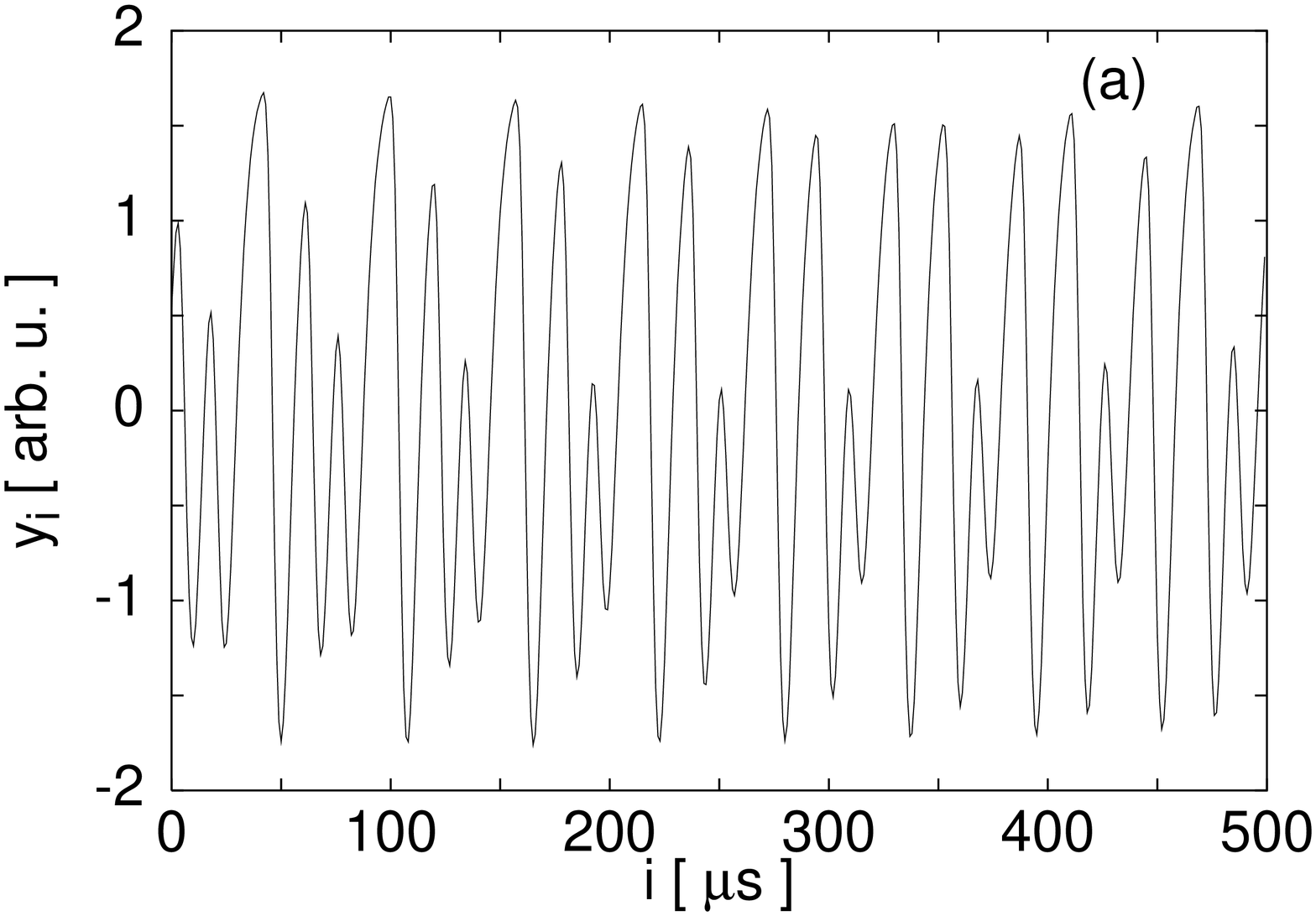,width=8cm,angle=270}
\hglue5mm\psfig{file=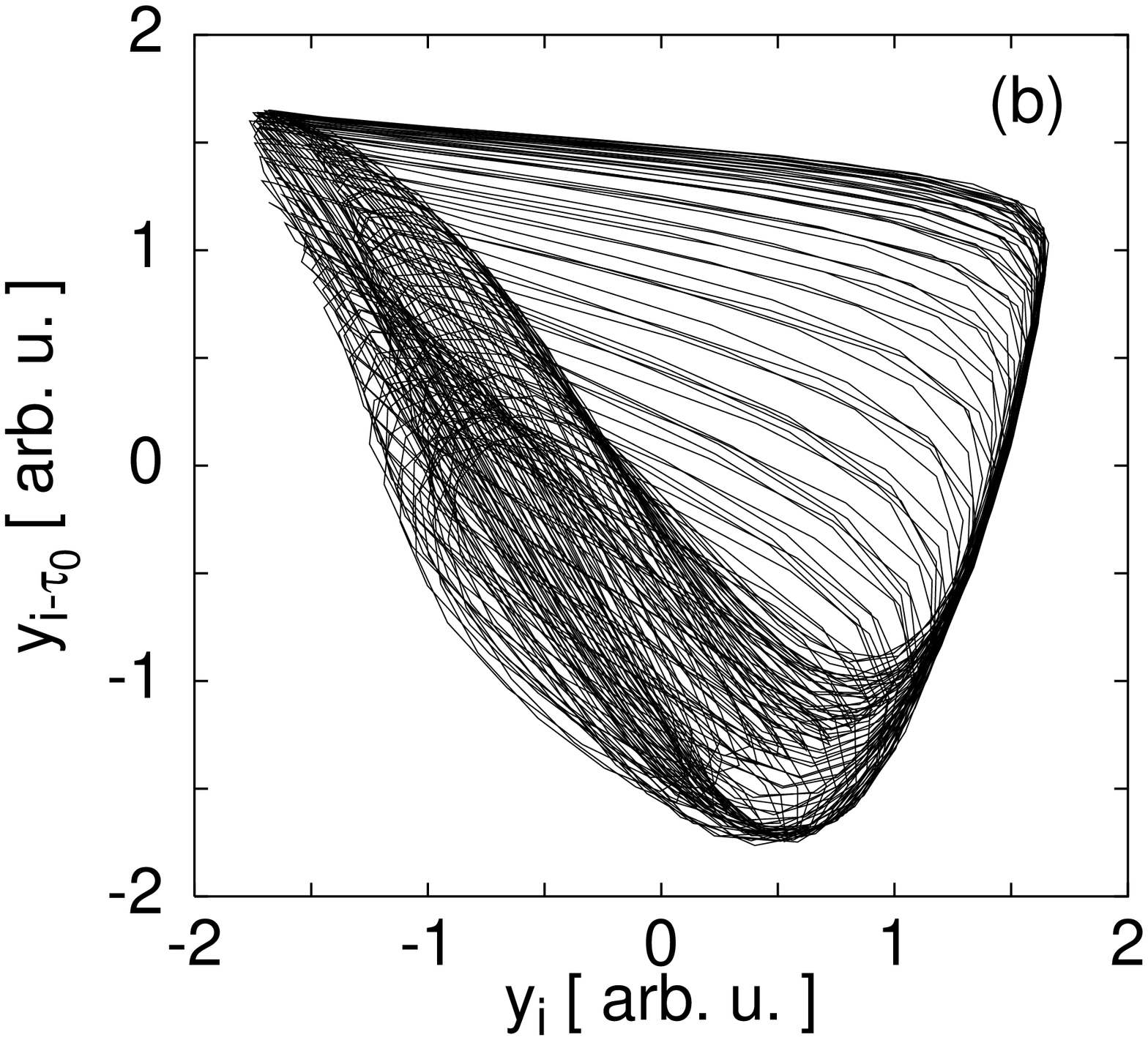,width=6.5cm,angle=270}}
\caption{\protect\small First 500 data of set with $\tau_0=50$ and
the first 2000 points in a two dimensional delay representation.}
\label{fig.r50-ts}
\end{figure}

The left panel of Fig.~\ref{fig.r50-ts} shows the first 500 points of 
the time series. The delay time evidently represents a relevant scale of the 
system, since it corresponds to an approximate periodicity of the signal.

\begin{figure}[ht]
\centerline{\psfig{file=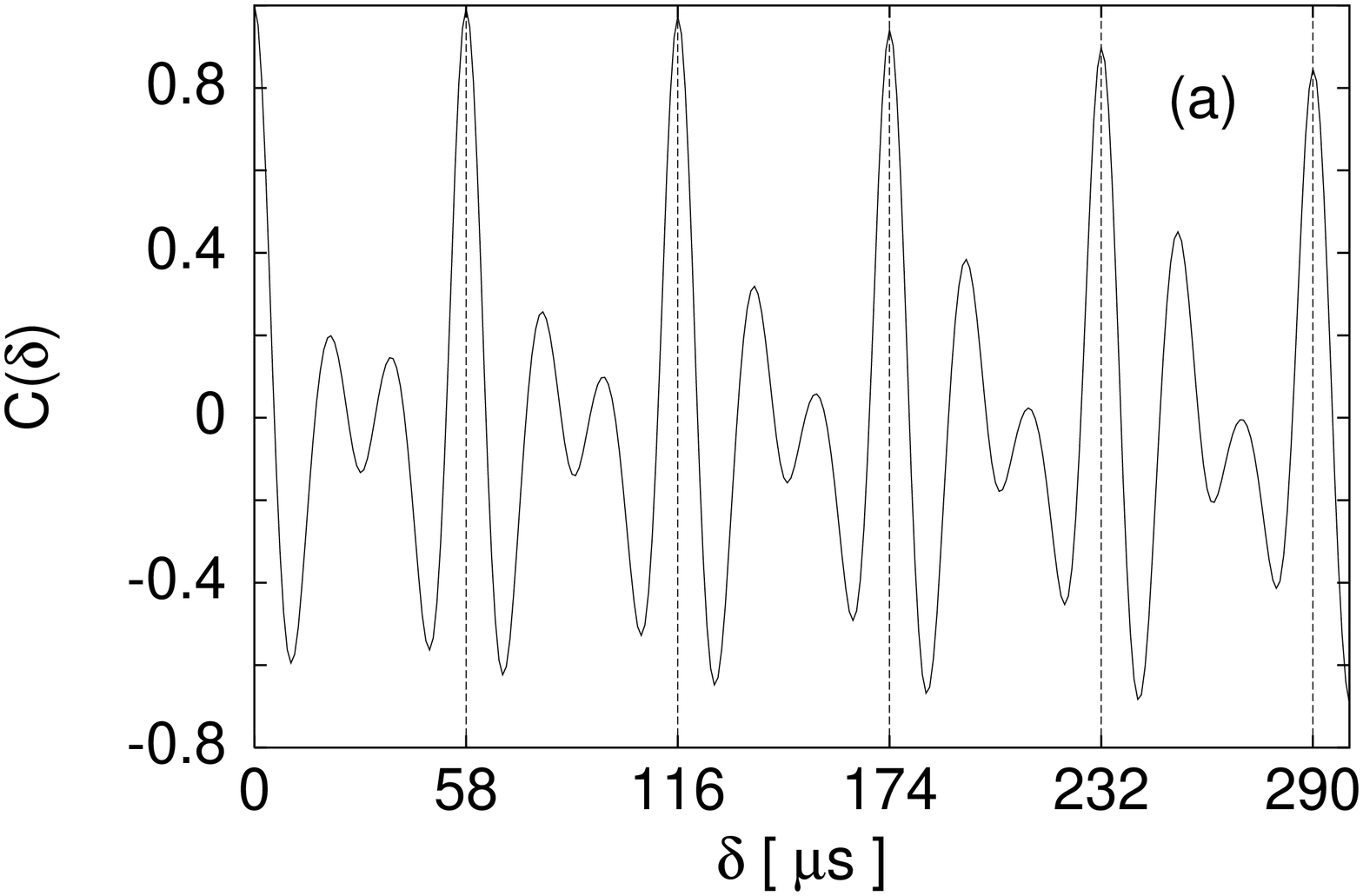,width=7.5cm,angle=270}
\psfig{file=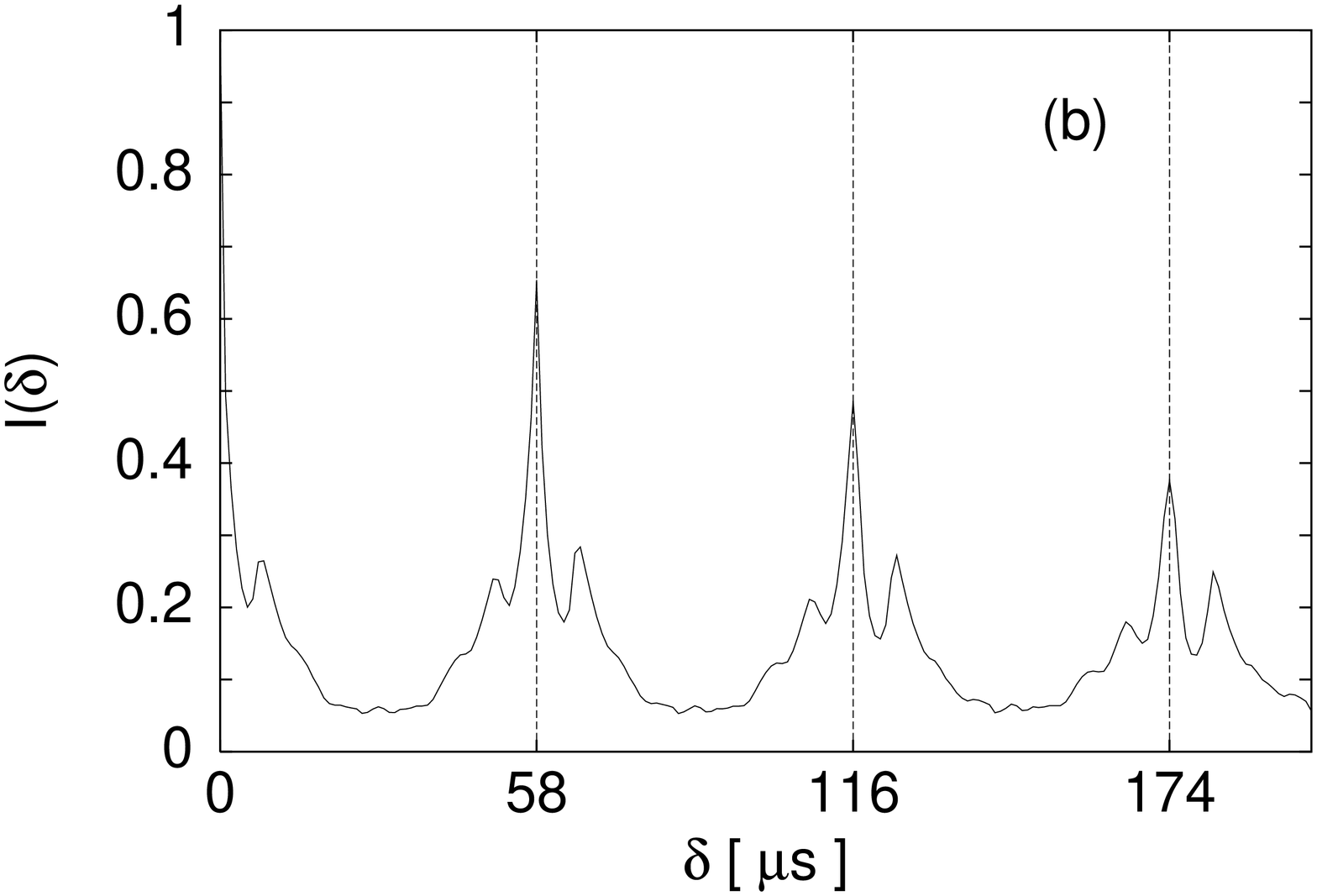,width=7.5cm,angle=270}}
\caption{\protect\small Autocorrelation (left panel) and time delayed mutual
information (right panel).}
\label{fig.r50-cor-mut}
\end{figure}

The right panel shows a projection on the two-dimensional state space 
$(y(t-\tau_0),y(t))$ (here, we use the experimental knowledge of 
$\tau_0$; later, the delay time will be estimated from the time series). 
It gives a reasonable insight about the overall shape of the hypersurface
on which the data lie. 
A characteristic signature of time-delayed systems is often
found in the behaviour of the autocorrelation function and of the mutual 
information which exhibit a marked peak for a time difference slightly
larger than the delay-time (see, for instance, \cite{AGLM92}). Such revivals
are clearly observable also in Fig.~\ref{fig.r50-cor-mut}. In both cases,
the first revival is situated at $\delta=58$, which is significantly larger 
than the delay time $\tau_0=50$. The reason is that the position of
the peak does not depend only on the delay time, but also on the response
of the system to the delayed feedback, so that one has to add an extra-delay
due to the response time \cite{Giacomelli99}. 

\begin{figure}[ht]
\centerline{\psfig{file=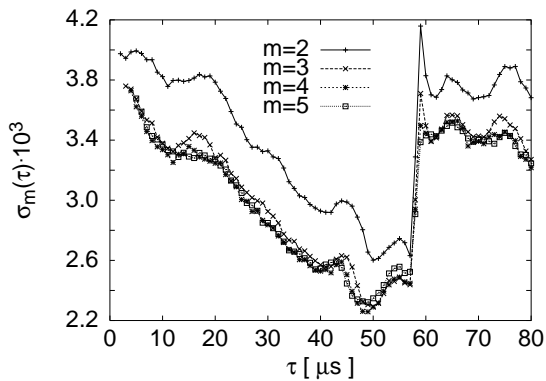,width=7.5cm,angle=270}}
\caption{\protect\small Relative forecast error (multiplied by a
factor of $10^3$) as a function of
$\tau$ for embedding dimensions $m=2,\ldots,5$.}
\label{fig.r50-fce}
\end{figure}

Now, we discuss the implementation of the embedding technique described in 
\cite{Buenner99}. In doing that, we consider that the delay $\tau_0$ is unknown
and thus it is one of the quantities to be determined. We start computing the 
one-step prediction error $\sigma_m(\tau)$ (normalized to the standard 
deviation of the data) as a function of the time $\tau$ used in the 
reconstruction (see Eq.~(\ref{eq:ourspace})) for $m=1,\ldots,5$.
For a fair comparison of the results for different embedding dimensions, we 
ensured that the average neighborhood size employed in the local fits 
was the same (within small fluctuations) for all $m$. This means that we 
use the full data set of length $10^6$ only for $m=5$, while shorter segments
are considered for smaller $m$. As a result, the systematic error due to the
linear ansatz is the same for all dimensions. The results for $m=2,\ldots,5$ 
are reported in Fig.~\ref{fig.r50-fce}, where a clear minimum is observed in 
all cases: for $m=2$, the minimum is at $\tau=51\pm1$, while for $m=3,4,5$,
$\tau=50\pm1$. Although the estimate of the delay time is in good agreement
with our expectations, the forecast error is not very sensitive to variations
in $\tau$ as, at most, it doubles when $\tau$ is grossly different from
$\tau_0$. The only exception is the case $m=1$, not reported in the figure
since it is approximately 50 times larger.

For $m \geq 2$ the minimal forecast error does never decrease below 
$ \sigma \approx 2.3\cdot10^{-3}$, which is mainly due to to the
dynamical noise  
present in the experiment(see also Sec.~\ref{sec.exp-setup}). In fact, the
minimum error does not change by independently varying the size of the 
neighbourhood of the fit and the sampling time; therefore we can conclude 
that it can neither be attributed to modelling errors nor to 
the approximations inherent the map-ansatz (i.e. the projection of the
infinite dimensional phase-space onto a finite-dimensional one).
In Sec.~\ref{sec.r150-400}, we shall find the same limitation on the FCE
also for larger values of $\tau_0$.

\begin{figure}[ht]
\centerline{\psfig{file=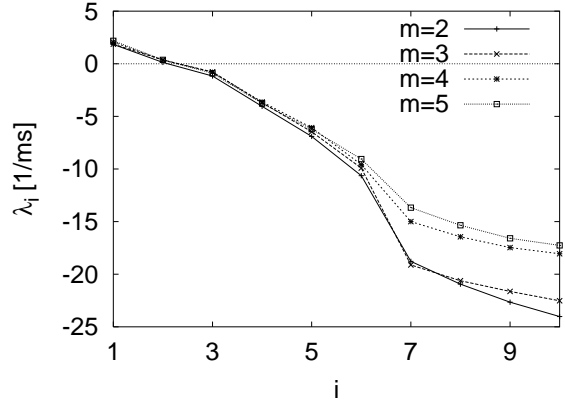,width=8cm,angle=270}}
\caption{Lyapunov spectrum of the $\tau_0=50$ data set. The first
exponent is estimated to be $\lambda_1\approx 1.8\ [1/{\rm ms}]$, while the
second one is about a factor 10 smaller.}
\label{fig.r50-lyaps}
\end{figure}

We now proceed with the determination of the Lyapunov spectrum by iterating 
the model equations.  For this estimate we use again a variable number of 
points (depending on $m$) in order to ensure the
same neighborhood size for the various embedding dimensions. The results are 
reported in Fig.~\ref{fig.r50-lyaps} (Please note that the Lyapunov exponents
are expressed in $1/{\rm ms}$ units rather than in the sampling frequency 
$1/\mu s$.). This analysis reveals only one 
positive Lyapunov exponent. One can also see that the first six exponents
agree for all embedding dimensions, while irregular deviations are
observed for the smaller ones. The reason for such deviations is not clear. 
It does not seem to be a systematic error, since there is no systematics 
observable by changing $m$. A possible explanation is a sensitivity of the
very negative Lyapunov exponents to the details of the model which certainly 
change upon varying $m$ (because of the noise). In any case, 
the exponents involved in the determination of the Kaplan--Yorke dimension 
are not influenced and we find the following estimates for the
dimension: $3.27$, $3.42$, $3.46$, $3.56$ for $m=2,3,4$ and $5$, 
respectively. By averaging these values, we can conclude that 
$D_{KY}=3.4\pm 0.1$. This value is small enough to allow us applying the 
methods of conventional nonlinear time-series analysis to validate the 
results. All numerical tools used here for this
purpose are part of the TISEAN package~\cite{Tisean99,Tisean}.

\begin{figure}[ht]
\centerline{\psfig{file=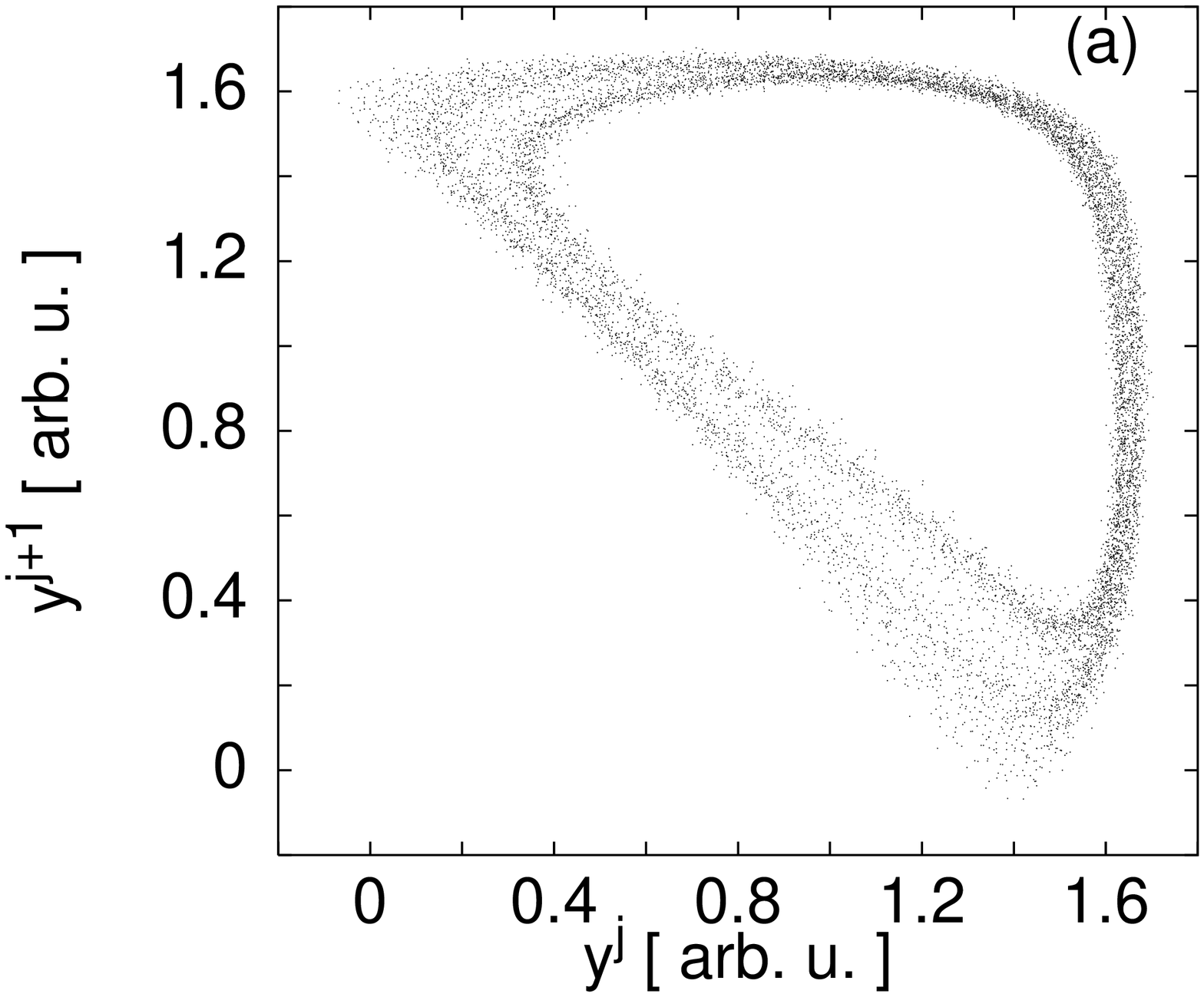,width=7cm,angle=270}
\psfig{file=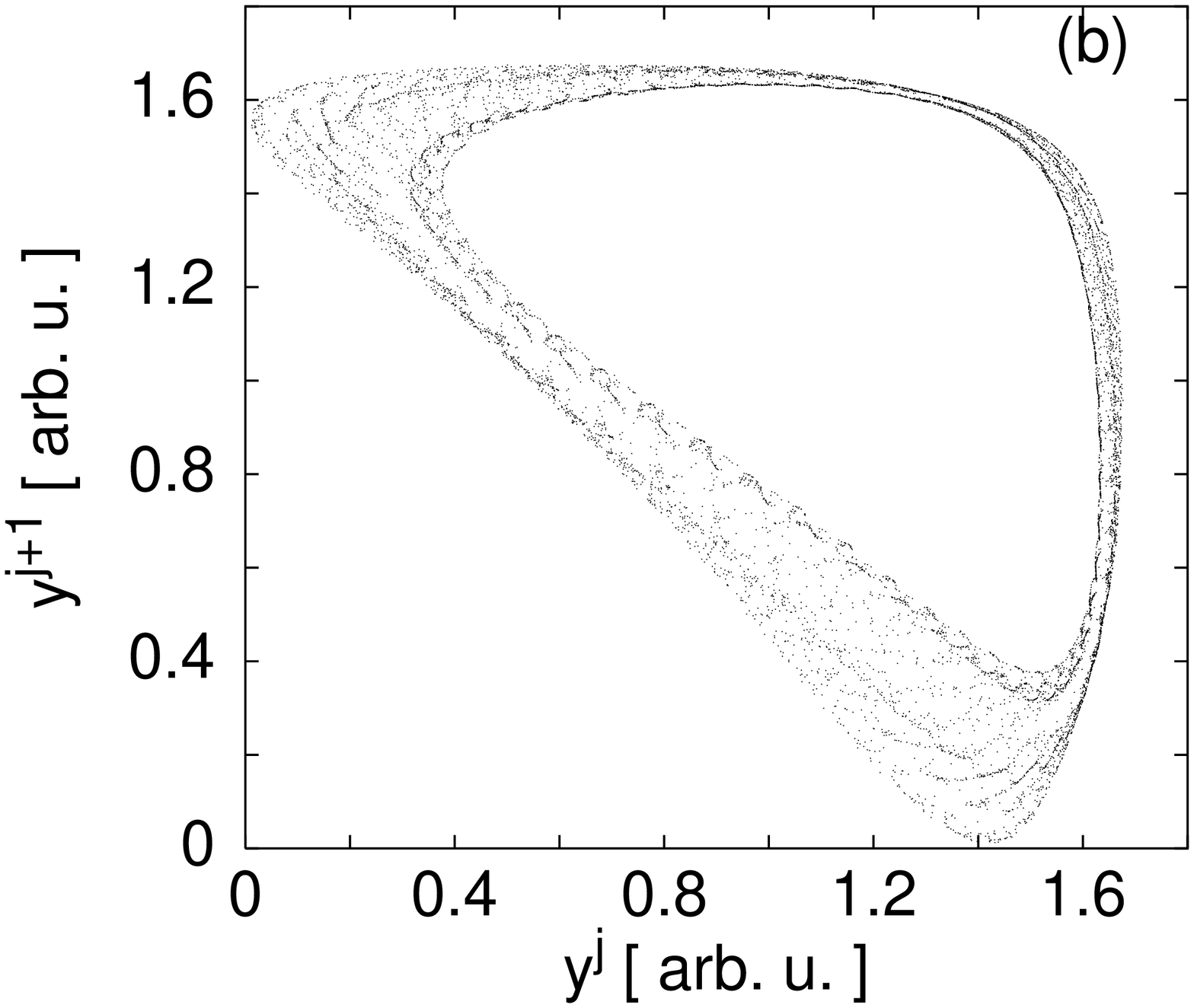,width=7cm,angle=270}}
\caption{Poincar\'e section of the flow data. Left panel before, right
panel after applying a noise reduction scheme.}
\label{fig.r50-poincare}
\end{figure}

Since the marginally stable direction of flow data is unfavorable for
the nonlinear analysis, we remove it. We construct a Poincar\'e
map in the hyper-plane $\dot{y}=0$ by collecting all local maxima. 
The resulting data is shown in panel (a) of
fig.~\ref{fig.r50-poincare}. Although the noise level of the original
flow data is fairly low, the Poincar\'e section is rather noisy. We
thus apply the noise reduction algorithm presented in~\cite{Grassberger93a}.
The result after 10 iterations (which turned out to be sufficient) is
shown in panel (b) of Fig.~\ref{fig.r50-poincare}.

The figure shows that the noise reduction scheme allows resolving
some singularities in the invariant measure of the data. Since the first 
negative Lyapunov exponent is close to zero (see Fig.~\ref{fig.r50-lyaps}), 
even the data of the Poincar\'e map look very much like flow data.
Moreover, a clustering (a kind of phase locking) of the points is clearly 
visible: this is an artifact of the noise reduction which typically appears 
for flow data that are sampled almost in phase with the dominating frequency.
The following results, however, are not significantly influenced by this
clustering effect.

\begin{figure}[ht]
\centerline{\psfig{file=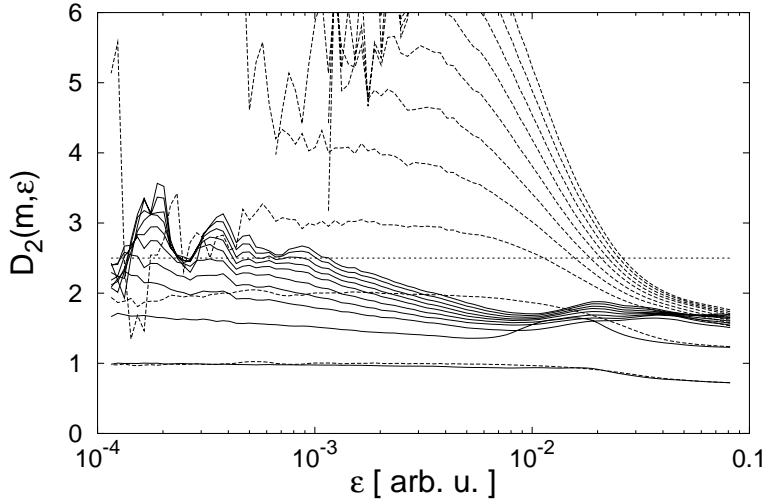,width=10cm,angle=270}}
\caption{\protect{$D_2$} for the data in the Poincar\'e section. 
The solid lines show the result for the data after noise reduction, 
the dashed ones for the original data. The dashed horizontal line
corresponds to the Lyapunov dimension obtained from the DM-model.}
\label{fig.r50-d2}
\end{figure}

In Fig.~\ref{fig.r50-d2}, we show the behaviour of the correlation 
dimension~\cite{Grassberger83d}. The analysis was done for (standard) 
embedding dimensions $d_E=1,\ldots,10$.  The dashed lines stem from the
data before noise reduction (from bottom to top with increasing
$d_E$). Due to the dominance of the noise, the only plateaus one sees are 
those corresponding to the embedding-space dimension, $d_E$. The solid 
lines in the figure show the behaviour after noise reduction. In this case, 
there is a nontrivial scaling behaviour on small length scales (smaller
than $10^{-3}$). Though the estimates are still too fuzzy to derive a well
defined dimension, the behaviour is in agreement with the Kaplan-Yorke
dimension which corresponds to the horizontal line.

In the reconstructed phase space, one can obtain a model-free estimate of 
the maximal Lyapunov exponent by following the divergence of
nearby segments of the trajectory. Also in this case (by implementing the
procedure described in Ref.~\cite{Kantz94}), while we cannot draw a
definite conclusion about the numerical value of the Lyapunov exponent,
we can certainly say that the growth rate is consistent with the
previous estimate of the Lyapunov spectrum.

The FCE is a local measure of the validity of a given model. However, as 
already seen in \cite{Buenner99}, a small FCE is a necessary condition
for a global reproduction of the observed dynamics, but it is not at
all sufficient. For this reason we have also decided to iterate the
optimal models obtained for each value of $m$, starting from meaningful 
initial conditions (i.e. on the attractor). The ($m=1$)-dynamics leads to 
unbounded solutions. Accordingly, the hypothesis that a scalar model provides
a convincing reconstruction of the dynamics can be rejected (notice that this
conclusion was transparent already in view of the very large FCE found 
for $m=1$). For $m \ge 2$, the forward iterates of the local linear models 
remain bounded and their two-dimensional representations appear to be very 
similar to the experimental data (see Fig.\ref{fig.r50-ts}), so that
we can conclude that the approximations implicit in the reconstruction
with $m_d=2$ are small and harmless. In the second section, we mentioned that 
the minimal realistic model for a $CO_2$ laser with feedback involves 6 
variables: the above result suggests that 4 of them can be, in a sense,
adiabatically eliminated. This conclusion goes even beyond the application
of the center manifold technique that allowed reducing the initial 
6-dimensional model (in the case of short delay) to a 3-dimensional 
one \cite{Varone95}.

Finally, we tried to model the dynamics by fitting autonomous models in 
the conventional embedding spaces. 
Although the attractor is fairly low dimensional, these models failed
by either converging to fixed point solutions or diverging to infinity. 
Therefore, we can summarize the discussion on the low-dimensional chaotic 
dynamics by stating that the construction of a DM-model with $m\ge 2$ 
components is not only consistent with the application of standard methods 
of nonlinear time series analysis, but provides a more stable modellization.
The superiority of the DM approach will become more transparent in the
next section where we apply the method to high-dimensional signals, a
case in which Takens-like embedding necessarily fails to give meaningful
answers.

\section{The high dimensional case}
\label{sec.r150-400}
Past experience and the analogy with spatially extended systems indicate that 
large time-delays in the feedback loop enhance the complexity of the 
dynamics. In this section, we analyse the data sets obtained for 
$\tau_0=150$ and $\tau_0=400$, respectively. It will turn out that the
attractor dimensions $D_f$ of these two series are indeed quite large, so
that the standard embedding would fail due to the requirement $m>2D_f$. 
The aims are (i) to show that our ansatz allows us to model the data with 
reasonable accuracy and (ii) to estimate the minimal embedding dimension 
$m_d$ for a meaningful reconstruction. This gives an upper bound for the 
number $d$ of variables involved in the system \cite{Buenner99}.

As in the previous section, we employ the forecast error to identify the 
optimal delay time $\tau_0$. Again the length of the time series is varied 
from 20000 for $m=2$ to $10^6$ for $m=5$ to ensure a neighbourhood size
almost independent of $m$ (approximately one percent of the attractor size).

\begin{figure}[ht]
\centerline{\psfig{file=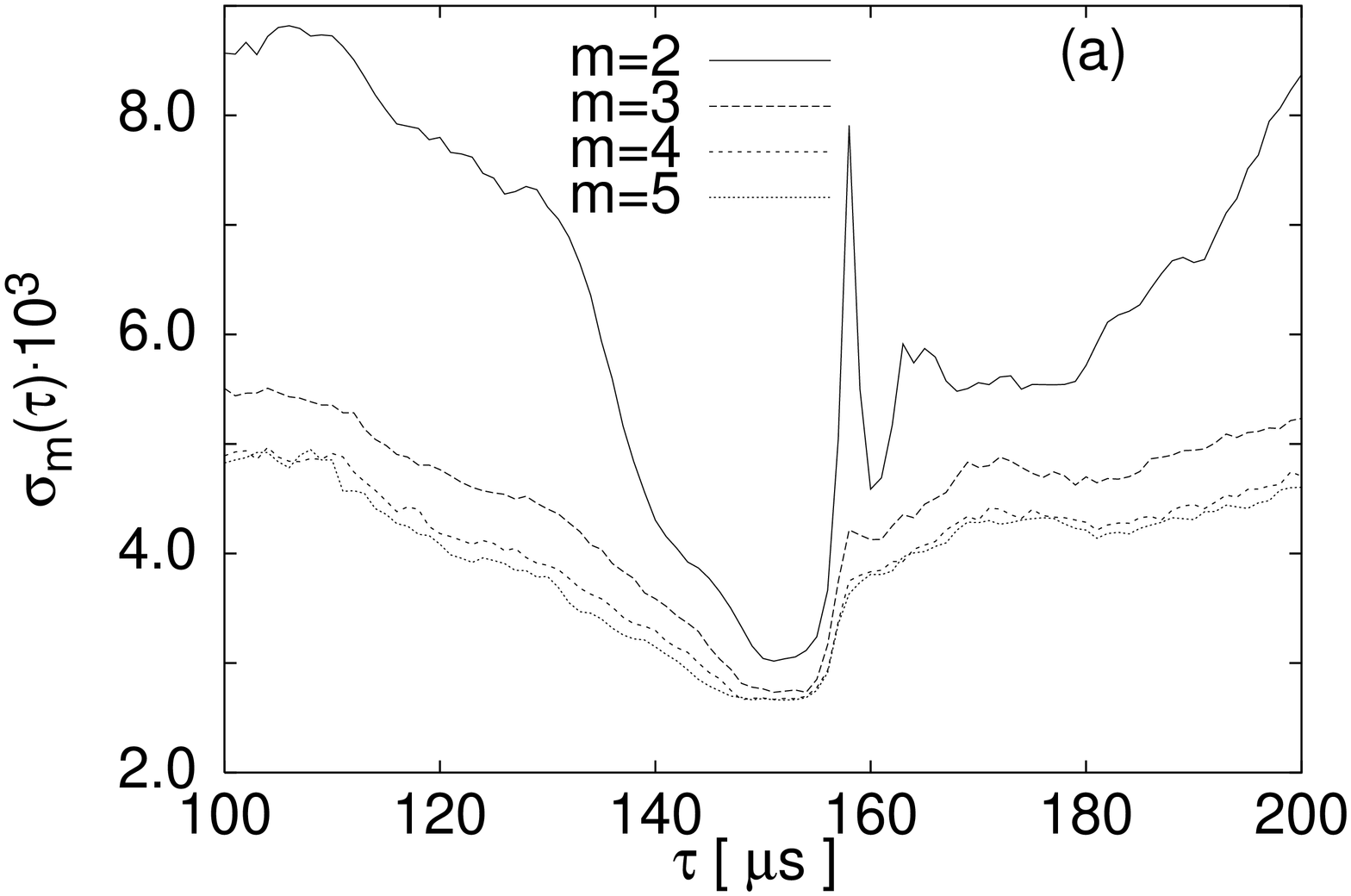,width=8cm,angle=270}
\psfig{file=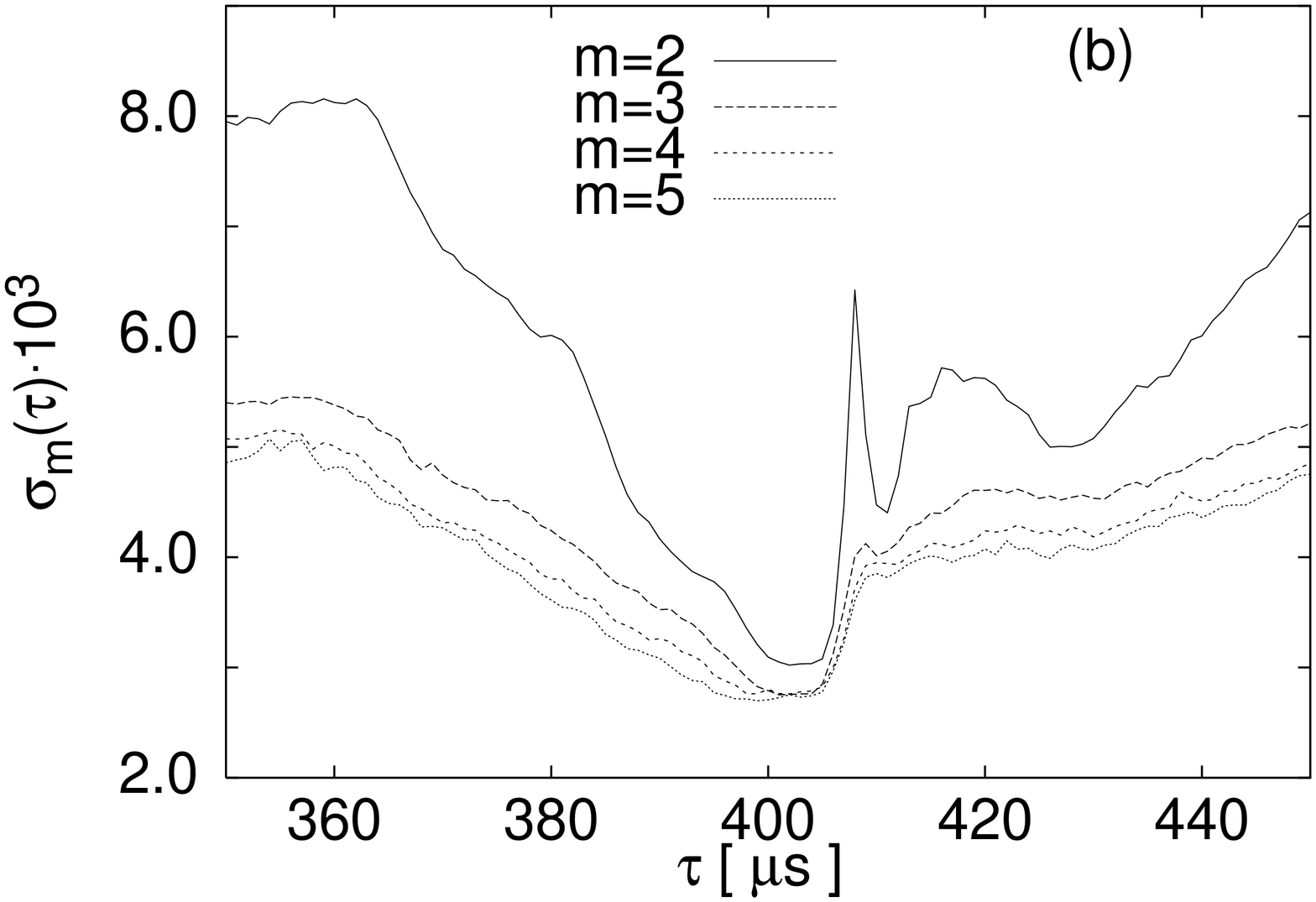,width=8cm,angle=270}}
\caption{Relative one-step prediction error (multiplied by $10^3$ in both
panels) for the data sets with $\tau_0=150$
(left panel) and $\tau_0=400$ (right panel), respectively. The
different curves correspond to different embedding dimensions.}
\label{fig.r150-r400-fce}
\end{figure}

The behaviour displayed by the FCE in Fig.~\ref{fig.r150-r400-fce} is 
qualitatively similar to that observed in the low--dimensional regime 
(reported in Fig.~\ref{fig.r50-fce}). At variance with the previous case, 
now there is a significant error reduction in passing from $m=2$ to $m=3$, 
but the minimum error does not further decrease by increasing $m$ beyond 3, 
meaning that the limit imposed by the noise level has been reached. 
Accordingly, this indicator does not allow drawing a definite conclusion 
about the relevance of additional degrees of freedom besides the first 3-ones. 
A further difference with the previous case is the larger value of the
FCE for $\tau$ significantly different from $\tau_0$. In that region,
the information contained in the second delayed window becomes increasingly 
negligible because of the decay of temporal correlations, so that the
performance of our technique does not differ from that of the standard
embedding approach for the same value of $m$. As here, at variance with 
the previous section, the attractor dimension is definitely larger than $m$ 
itself, larger FCE's have to be expected for low-$m$ models. 

The estimates of the delay times are reported in 
table~\ref{tab.r150-r400-tau}. Each value corresponds to the centre of the 
dip, while the error is the half-width of the dip. As in the low-dimensional 
case, we have investigated the peaks of the time delayed mutual information 
and of the autocorrelation. We find again an offset with respect to
the known delay. For the data set corresponding to $\tau_0=150$, the peak 
lies at $\tau=157$ while for $\tau_0=400$ the peak is found  at $\tau=407$. 
All such data together indicate that the offset is essentially independent
of the delay, confirming that it is just the response time of the laser
system to the feedback. 

\begin{table}[ht]
\begin{center}
\[
\begin{array}[t]{|c|c|c|}
\hline
m & \tau_0=150 & \tau_0=400\\
\cline{1-3}
2 & 152\pm 2 & 402\pm 2 \\
3 & 152\pm 2 & 402\pm 3\\
4 & 151\pm 3 & 401\pm 4\\
5 & 151\pm 3 & 401\pm 4\\
\hline
\end{array}
\]
\end{center}
\caption{Estimates of the delay times for the $\tau_0=150$ and the
$\tau_0=400$ data sets, respectively.}
\label{tab.r150-r400-tau}
\end{table}

Together with the delay time, we now want to estimate $m_d$. Again, 
we require that a good model does not only yield good
one-step forecasts but reproduces all the properties of the
experimental data, when long trajectories are created by iterating the
model itself. In other words we propose to compare global properties of 
the data sets~\cite{Buenner99}. Once we have verified that the trajectories 
do not escape to infinity under iteration of the model, we have determined
the histograms of the data (i.e. the one-dimensional projection of 
the invariant measure) and the power spectra. Furthermore, we have 
investigated the convergence of the Lyapunov spectra as a function of $m$
and the behaviour of the cross--prediction errors between numerically
generated time series and experimental data.
For the sake of completeness, we have also investigated other indicators
such as the mutual information, but we do not comment on the corresponding
results as they simply confirm the overall scenario.

\begin{figure}[ht]
\centerline{\psfig{file=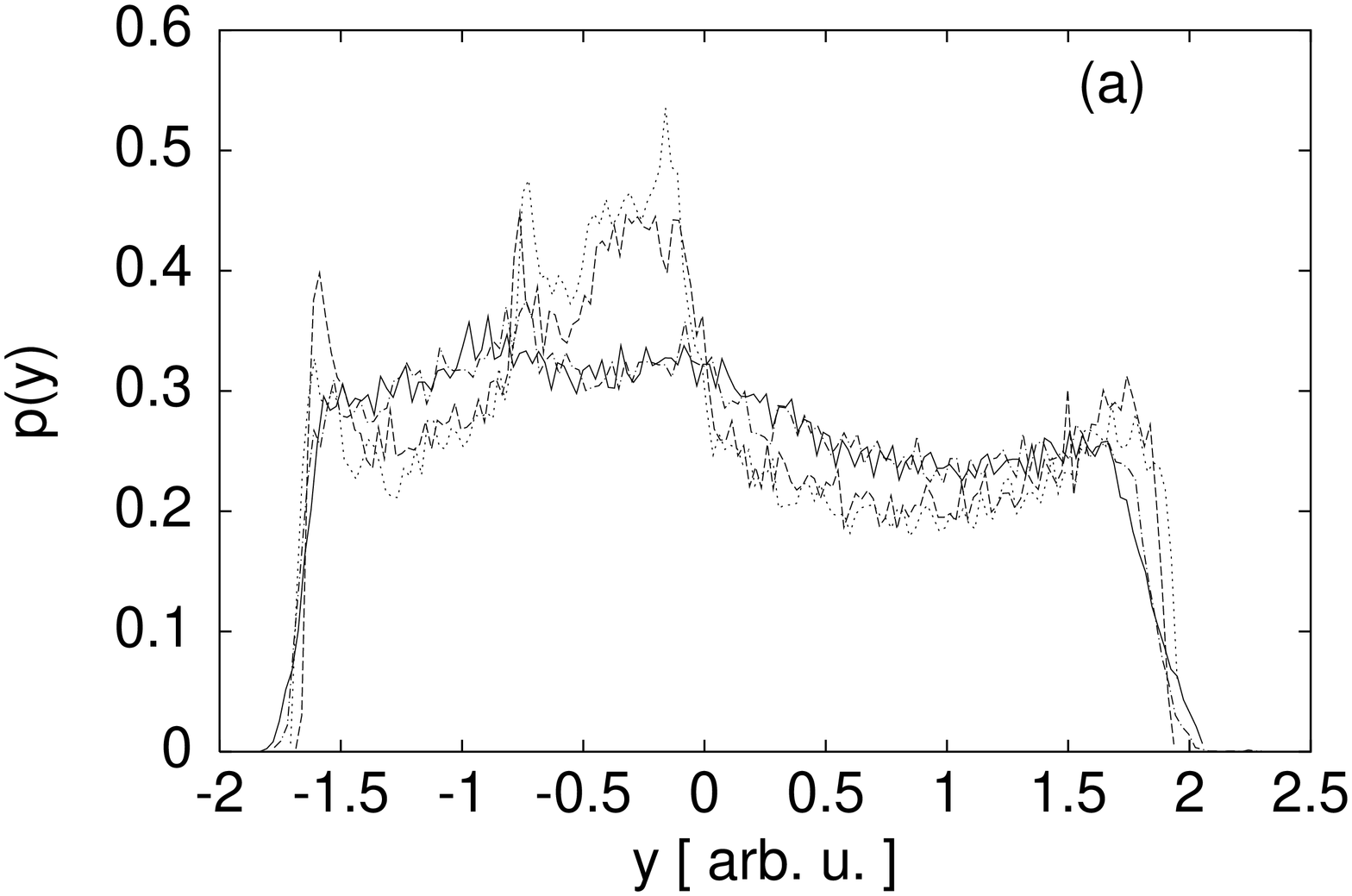,width=8cm,angle=270}
\psfig{file=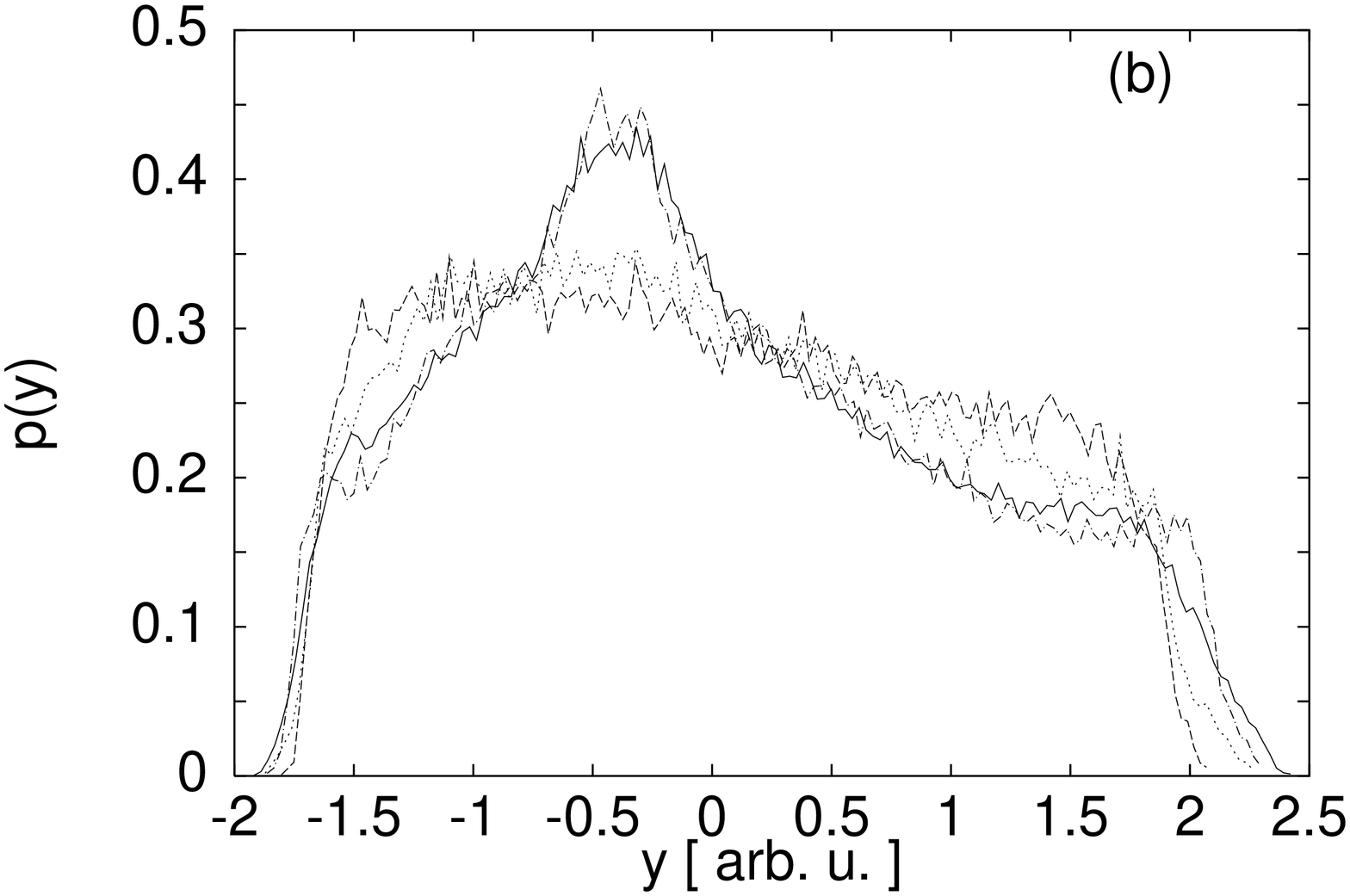,width=8cm,angle=270}}
\caption{One dimensional projection of the invariant measure of the
original data and for modelled data in different embedding
dimensions. Left panel (a): data set with $\tau_0=150$, right panel (b): data
set with $\tau_0=400$. In both panels, the solid line corresponds to
the original data, the dashed line to the fit for $m=3$, the dotted
one to $m=4$ and the dashed--dotted one to $m=5$.}
\label{fig.r150-r400-his}
\end{figure}

\begin{figure}[ht]
\centerline{\psfig{file=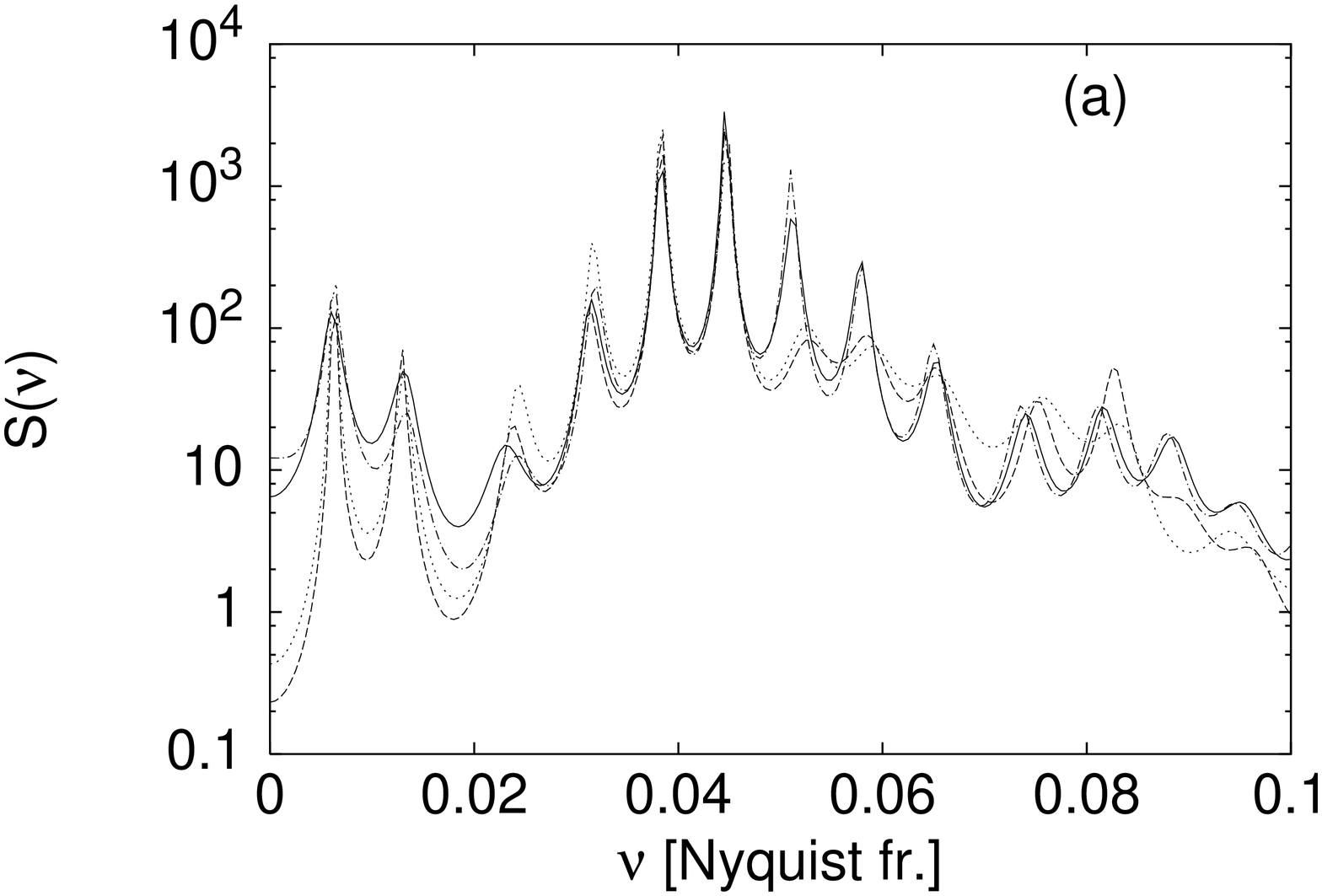,width=8cm,angle=270}
\psfig{file=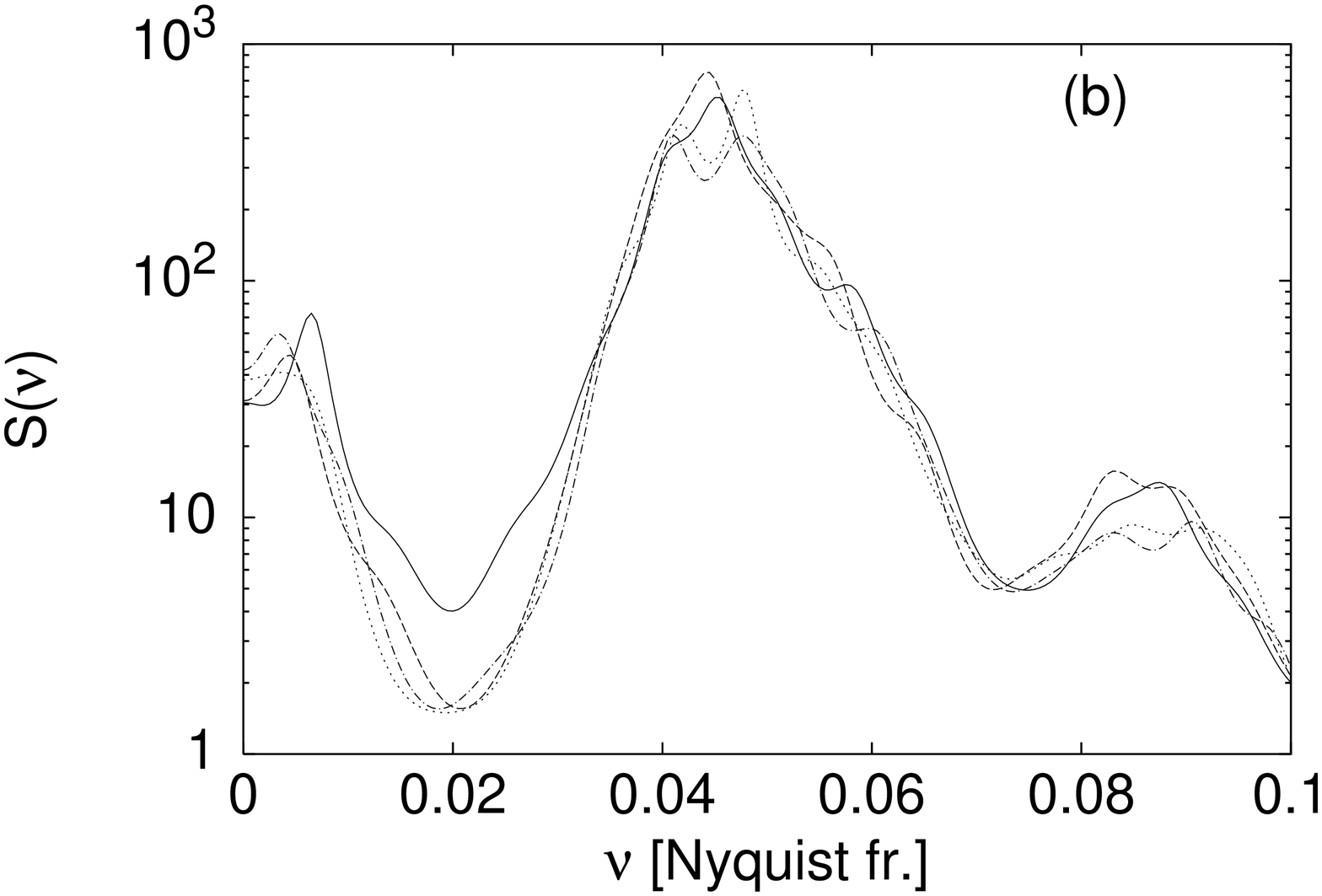,width=8cm,angle=270}}
\caption{Power spectrum of the
original data and for modelled data in different embedding
dimensions. Left panel (a): data set with $\tau_0=150$, right panel
(b): data set with $\tau_0=400$. The definition of the lines is the
same as in fig.~\protect{\ref{fig.r150-r400-his}}.}
\label{fig.r150-r400-spec}
\end{figure}

The comparison is performed as follows: for each embedding dimension we 
choose the optimal delay time from table~\ref{tab.r150-r400-tau} and 
accordingly generate an artificial trajectory which is used to compute 
the quantities to be compared with the corresponding ones obtained from
the original data. If all of them (histograms, power spectra and 
Lyapunov exponents) agree for a given $m$, we can conclude that the $m$-value
allows a faithful reconstruction. The smallest such $m$ is then defined to 
be $m_d$. Figures~\ref{fig.r150-r400-his} and
\ref{fig.r150-r400-spec} show the results for the histograms and the 
power spectra, respectively.

The histograms reported in Fig.~\ref{fig.r150-r400-his} show a nonnegligible
dependence on the embedding dimension (the curve for $m=2$ is not shown as 
it deviates even more strongly from the expected distribution). It is quite
curious to see that while for $\tau_0 = 150$, simpler models ($m=3$, 4) 
give rise to a (spurious) peak around $y= - 0.5$ which disappears upon
further increasing $m$, the opposite is observed for $\tau_0 = 400$.
We can understand this phenomenon by first noticing that $y \approx -0.5$
corresponds to an unstable fixed point of the dynamics (corresponding to the 
steady lasing state). Therefore, it turns out that a correct estimation
of the stability of the fixed point is certainly crucial for a correct 
determination of
the probability density in its surroundings. A posteriori, we can conclude
that ``underembedding'' leads to a wrong estimation of the local stability.

The results for the power spectra are less clear. While for $\tau_0=400$,
the relevant structures of the spectra are reproduced sufficiently well already
for $m=3$, for $\tau_0=150$, differences are still observed by comparing the
cases $m=3$ and $4$. Anyway, the relevant frequences are reproduced already
for $m=3$, so that a clear conclusion about the appropriate $m_d$-value is 
quite hard to draw.

A more global test of the validity of a reconstructed model is through the 
computation of the cross--prediction error which is a measure of the 
``distance'' between dynamical regimes. One might, for instance,  be willing
to compare different segments of the same time series to test the stationarity
of the process as in Ref.~\cite{Schreiber97}. In our context, the goal is
to compare the original experimental signal $X=\{x_n\}$ with the computer
generated trajectory $Y=\{y_n\}$, obtained by iterating the reconstructed 
model. More precisely, one uses the time series $Y$ as a data base
to make a prediction $\tilde{x}_{n+1}(Y)$ for each possible value of $n$. 
This can be done by identifying the closest $Y$-points to each $X$ point 
in the $(2m)$-dimensional state-space and using them to construct a local 
linear model (see Ref.~\cite{Buenner99}).  The
average cross--forecast error is then defined as
\be
\chi(X,Y)=\sqrt{\frac{\sum_{i=m}^N
\left(\tilde{x}_{n}(Y)-x_n\right)^2}{(N-m)\sigma^2(X)}}\;,
\label{eq.cross-def}
\ee
where $\sigma^2(X)$ is the variance of the time series $X$. Notice that
$\chi(X,X)$ reduces to the standard forecast error. Therefore, we expect 
$\chi(X,X)$ and $\chi(X,Y)$ to be approximately equal if $X$ and $Y$ 
follow the same dynamics.

\begin{table}[ht]
\begin{center}
\[
\begin{array}[t]{|c|c|c|}
\hline
m & \tau_0=150 & \tau_0=400\\
\cline{1-3}
\mbox{O} & 5.0\cdot 10^{-2} & 4.8\cdot 10^{-2} \\
\cline{1-3}
2 & 11.9\cdot 10^{-2} & 6.4\cdot 10^{-2}\\
3 & 11.9\cdot 10^{-2} & 6.0\cdot 10^{-2}\\
4 & 10.3\cdot 10^{-2} & 5.9\cdot 10^{-2}\\
5 & 5.9\cdot 10^{-2} & 6.1\cdot 10^{-2}\\
\hline
\end{array}
\]
\end{center}
\caption{Estimates of the cross--prediction errors for the
$\tau_0=150$ and the $\tau_0=400$ data sets, respectively. The row
indicated by O is the usual prediction error using the original time
series itself.}
\label{tab.r150-r400-cross}
\end{table}

The results are shown in Tab.~\ref{tab.r150-r400-cross}. The first row
shows the forecast error $\sigma(X,X)$ for both the $\tau_0=150$ and
the $\tau_0=400$ data set. The forecasts are made with a zeroth order
model (in five plus five dimensions) as described in 
Ref.~\cite{Schreiber97,Buenner99}. These results serve as references for the
next lines. There, the cross--forecast errors are shown for data
produced in the indicated embedding dimensions. Again the estimate of
the error is done in five plus five dimensions. As for the spectra,
we see differences  between the two data sets. While for the $\tau_0=150$
data there is an improvement if $m$ is increased from $4$ to $5$, no such 
trend is visible for the $\tau_0=400$ data. There, the cross--forecast error
is almost constant and comparably small already for $m=2$.

\begin{figure}[ht]
\centerline{\psfig{file=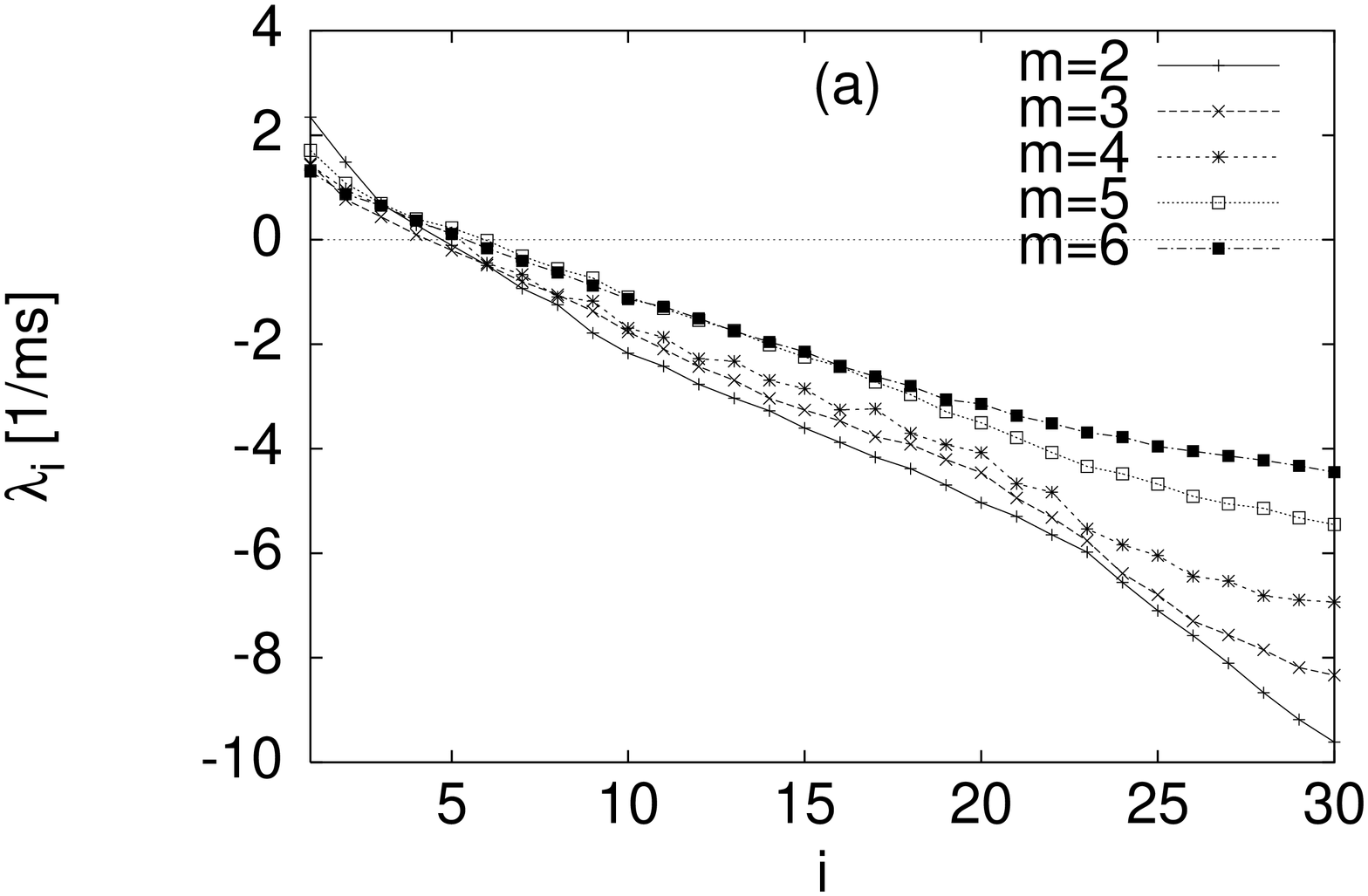,width=8cm,angle=270}
\psfig{file=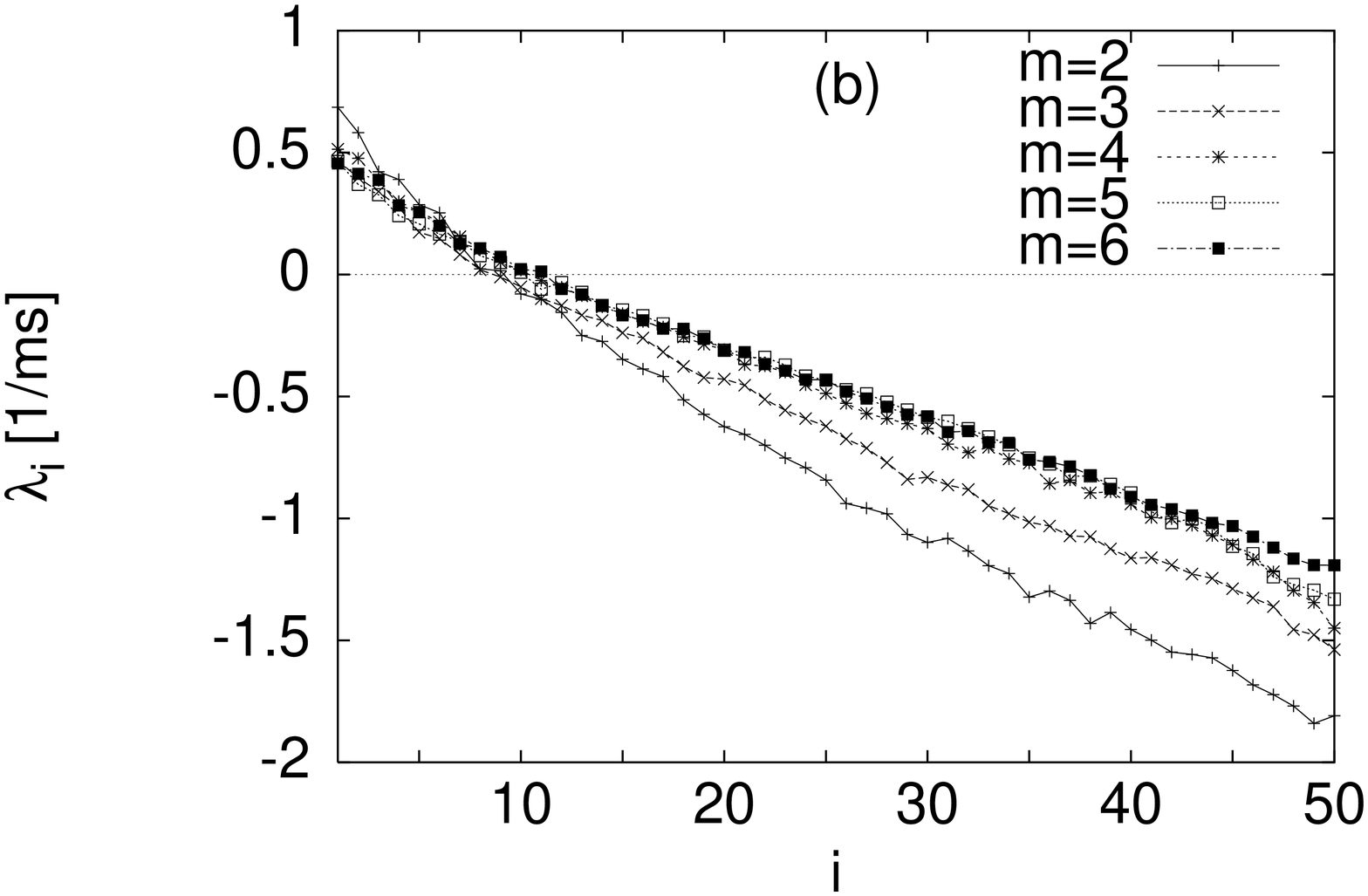,width=8cm,angle=270}}
\caption{Lyapunov spectra for the $\tau_l=150$ (panel a) and the
$\tau_0=400$ (panel b) data sets. Note that the exponents are given in
units of $1/ms$ and not in $1/\mu s$.}
\label{fig.r150-r400-lyaps}
\end{figure}

The convergence of the Lyapunov spectra as a function of $m$ is our last test. 
The resulting spectra are shown in Fig.~\ref{fig.r150-r400-lyaps}. The
left panel shows the first 30 exponents for $\tau_0=150$; in analogy with
the previous analysis, we see that a convincing convergence is achieved only 
for $m \ge 5$ (the differences between the spectra for $m=5$ and $6$
concern the smaller Lyapunov exponents and are thus not dynamically relevant).
The right panel of Fig.~\ref{fig.r150-r400-lyaps} shows the first 50 
exponents for $\tau_0=400$. Again a faster convergence is observed as 
suggested from the cross-prediction error, although the Lyapunov spectrum
appears to be a slightly more sensitive indicator as $m=4$ is the minimum
embedding dimension for a convincing convergence. 

\begin{figure}[ht]
\centerline{\psfig{file=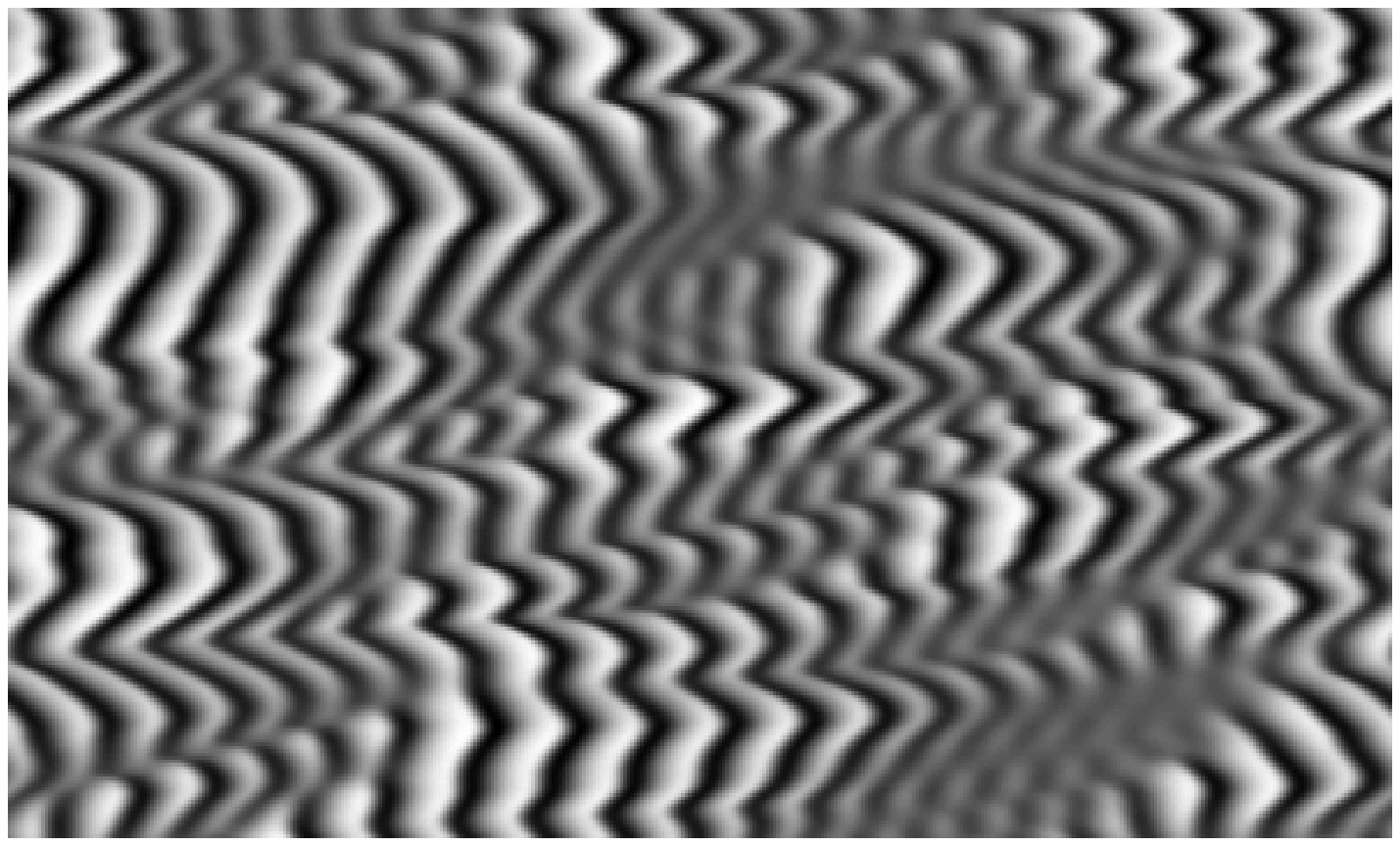,width=7cm}
\psfig{file=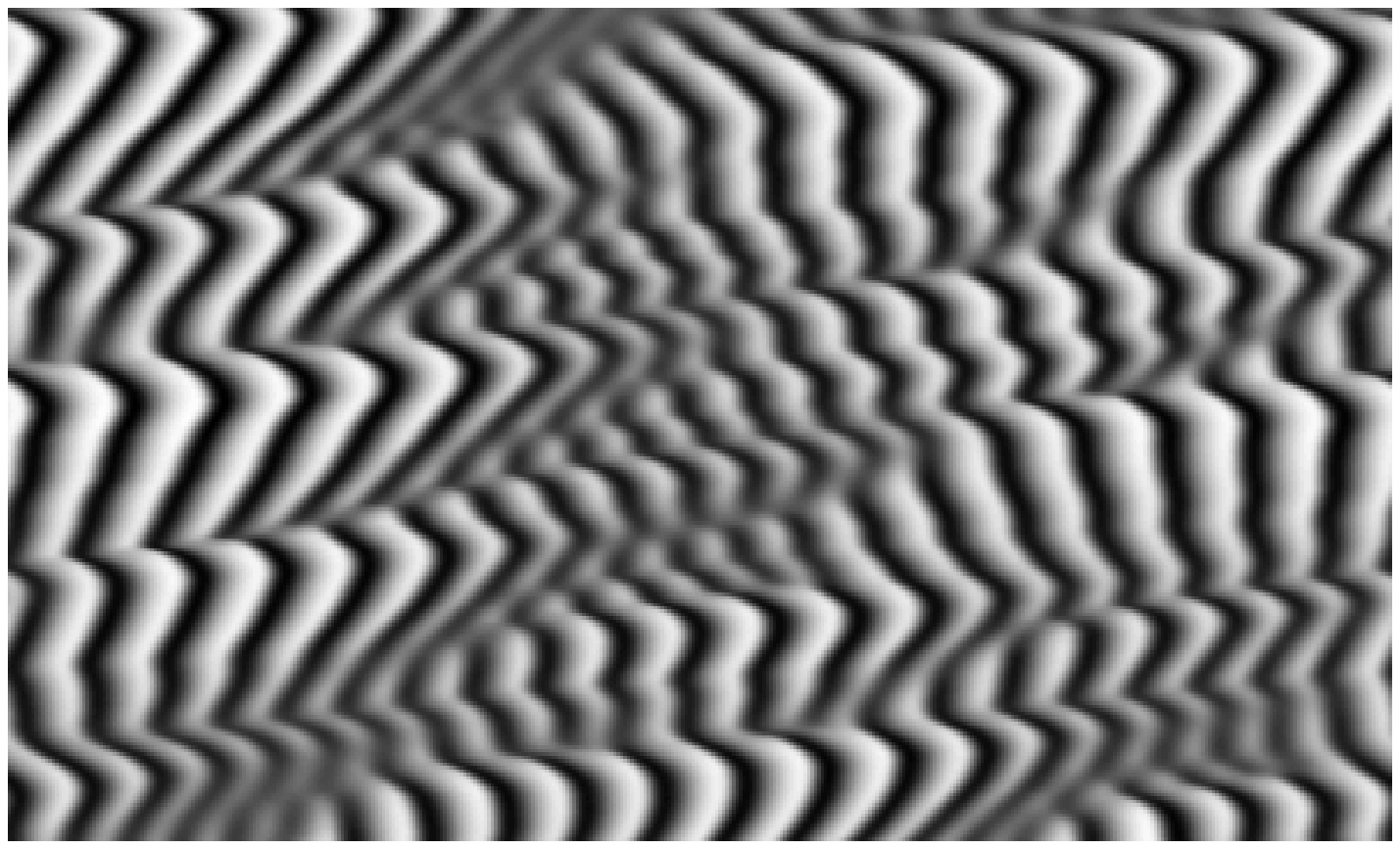,width=7cm}}
\caption{Space--time representation (as in
fig.~\protect{\ref{fig.r400-st}}) of the $\tau_0=400$ data set. The 
left panel shows the original data, the right panel the data obtained
by iterating a $m=3$ model.}
\label{fig.r400-space-time}
\end{figure}

Taking into account all the results presented so far, it is rather difficult 
to draw a final conclusion about $m_d$. It is certainly true that $m=5$ 
is always sufficient to guarantee a good agreement with the experimental data,
so that $m_d=5$ can be certainly considered as an upper bound to the minimal
number of effective degrees of freedom to be used to construct the state-space. 
On the other hand, some properties are already recovered for 
$m=3$. This is confirmed by looking at Fig.~\ref{fig.r400-space-time}, where
the space--time representation of the $\tau_0=400$ data set (left panel) 
can be compared with the computer generated trajectory (right panel) obtained
for $m=3$: no essential differences can be detected. While, it is not 
surprising that different indicators exhibit different sensitivity to 
modelling errors, it is somehow less understandable that the two data sets
 - corresponding to different delay times - give rise to different dependences 
on the embedding dimension. After a careful check of the original data,
we have found that the time series with $\tau_0 = 150$ (the one exhibiting
a slower convergence) is slightly corrupted by a noisy component at 50 Hz 
which evidently contributes to lowering the performance of our method. 
Anyway, in order to be on the safe side, in the remainder of the manuscript 
we shall always work with $m=5$.

The knowledge of the Lyapunov spectra allows to estimate both the
fractal dimension $D_{KY}$ through the Kaplan-Yorke formula and the 
Kolmogorov-Sinai entropy $h_{KS}$ through the Pesin relation. From the 
results reported in
Tab.~\ref{tab.r150-r400-lyaps}, we can see that the attractors are quite
high dimensional: in both cases, $D_{KY}$ is larger than the dimension of 
our state space and, more importantly, is definitely beyond the limit for 
a successful application of the standard embedding approach.

Furthermore, our results on the attractor dimension in the various regimes
provide the first experimental evidence that the dimension of a delayed 
system is proportional to the delay time. In fact, both in the low-dimensional
chaotic regime examined in the previous section and in the two
high-dimensional regimes studied here, we find approximately the same
dimension density $d = D_{KY}/\tau_0 \approx (64\pm 5) ms^{-1}$. This means that
the addition of $(16\pm 1) \mu s$ to the delay line contributes to 
increasing the dimension by one unit. The extensivity of the fractal
dimension, first noticed in \cite{Farmer82} when studying the Mackey-Glass
model, can be understood on the basis of the analogy with spatially extended 
systems which allows interpreting the delay as a ``spatial'' size of a 
suitable one-dimensional system \cite{GiPo96}.

\begin{table}[ht]
\begin{center}
\[
\begin{array}{|c|c|c|c|c|c|}
\cline{1-6}
\tau_0 & N & \lambda_{\rm max} $[1/{\rm ms}]$ & D_{KY} & h_{\rm KS}
$[1/{\rm ms}]$ & D_{KY}/\tau_0 $[1/{\rm ms}]$ \\ 
\cline{1-6}\cline{1-6}
50 & 1 & 2.5 & 3.4 & 2.5 & 68\\ \cline{1-6}
150 & 4 & 1.4 & 10.2 & 3.3 & 68\\ \cline{1-6}
400 & 11 & 0.55 & 23.8 & 3.0 & 59\\
\cline{1-6}
\end{array}
\]
\end{center}
\caption{Results exploring the Lyapunov spectra.
$N$ is the number of positive Lyapunov
exponents, $\lambda_{\rm max}$ the maximal exponent, $D_{KY}$ the
Kaplan--Yorke dimension and $h_{\rm KS}$ the Kolmogorov-Sinai.}
\label{tab.r150-r400-lyaps}
\end{table}

In contrast to the fractal dimension that is an extensive quantity,
the dynamical entropy $h_{KS}$ of delay feedback systems turns out to be 
independent of the delay. The results reported in 
Tab.~\ref{tab.r150-r400-lyaps} clearly confirm this expectation.
The reason for such a difference with truly extended systems (where the
dynamical entropy too is extensive) can be traced back to the units
of the ``time'' variable $\theta$ that have to be used for a meaningful 
space-time representation of delayed systems. As seen in 
Eq.~(\ref{eq.space-time}), $\theta$ is essentially equal to the actual
time except for a multiplicative factor equal to the delay time $\tau$
(for this discussion, the possible difference between $\tau_0$ and $\tau_1$ 
is immaterial).  Accordingly, a dynamical entropy measured in 
${\theta}^{-1}$ units has to be multiplied by $\tau$ and thus acquires an 
``extensive'' character as expected for a spatially extended system.
 
\begin{figure}[ht]
\centerline{\psfig{file=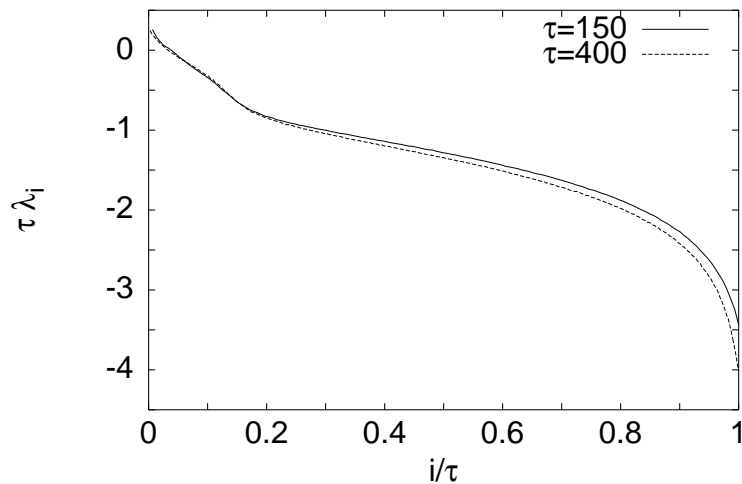,width=10cm,angle=270}}
\caption{Rescaled Lyapunov spectra for the $\tau_l=150$ and the
$\tau_l=400$ data sets. The exponents were calculated for $m=5$.}
\label{fig.r150-r400-scaled-lyaps}
\end{figure}

The extensive character of the dimension and the intensive nature of
the dynamical entropy both follow from a general property of the
Lyapunov spectrum that is invariant (in the limit of long delays) under
the simultaneous rescaling of the Lyapunov exponents by a factor $\tau$ 
and of the exponent's index by $1/\tau$ \cite{GiaLePo95}. Accordingly, the two
spectra reported in Fig.~\ref{fig.r150-r400-lyaps} should collapse onto
the same curve. This is indeed confirmed by 
Fig.~\ref{fig.r150-r400-scaled-lyaps}, which at the same time confirms
the correctness of the two-window embedding technique and provides
experimental evidence for the scaling behaviour of the Lyapunov spectrum.

The spectrum obtained for $\tau_0=50$ does not overlap equally well.
In complete analogy with space--time systems, this is certainly due to
finite-``size'' effects.

\section{Delay-time independence of the model}
\label{sec.cross_check}
We started developing our theory under the assumption that we deal
with delay differential equations such as in Eq.~\ref{dde-def.eq}. The success
of the two-window approach implies that the information contained in 
$\vv_n$ is equivalent to that in $(\vx(t),x_l(t-\tau_0))$ for $t=n\delta t$, 
provided that the correct value of $\tau=\tau_0$ is chosen in $\vv_n$. 
Accordingly, there exists a delayed map $y_{n+1}= g(\vv_n)$ whose dynamics is
equivalent to that stemming from the velocity field $f(\vx(t),x_l(t-\tau_0))$. 
As a consequence, as long as the velocity field does not depend 
explicitly on $\tau$ (as it is the case in our experimental set-up), 
the same holds true for the mapping $g$.

The outstanding consequence is that the model constructed from a data set 
with some $\tau_0=\tau_a$ can be used to simulate the behaviour for a 
different delay $\tau_0=\tau_b$. The only problem comes from the lack
of a global knowledge of $g$, which is inferred only in the region of the
state-space visited by the input data. Therefore, the extension of the model 
to a different delay  is possible only if, in the new regime, the dynamics 
does not leave the support of the invariant measure of the 
original one.

Iteration of a model constructed for $\tau_0=\tau_a$ with a different delay 
time $\tau_b$ can be executed in the following way. Let $\vv_i(\tau_a)$ be 
the vectors from the reference data set (our data base) and $\tvv_i(\tau_b)$ 
the vectors to be produced. In order to forecast $\tx_{n+1}$ we determine 
$g$ in a small neighborhood of $\tvv_n(\tau_b)$. This can be done by 
linearly approximating the set of all points $\vv_r(\tau_a)$ that are close 
to $\tvv_n(\tau_b)$. Next, we use the fitted law to iterate $\tvv_n(\tau_b)$. 
This is exactly the scheme of the cross-prediction described in the previous
section, with the only difference that here the past window of 
$\tvv_i(\tau_b)$ vectors differs from that of $\vv_i(\tau_a)$ ones.

More specifically, we have used the data set with $\tau_a=400$ to produce 
a trajectory simulating the behaviour for $\tau_b=150$. Since the 
dimension of the dynamics for $\tau_a$ is larger than that for $\tau_b$, we 
can expect that our artificial $\tau_b$-trajectory will not leave the 
region where we have to find $\tau_a$-neighbours (i.e., the support of 
the invariant measure for $\tau_b$ is a subset of that in the $\tau_a$ case). 
As an initial condition for the forecast, we used the first points of the
$\tau_a$ data, so that we have to discard the initial transient.

\begin{figure}[ht]
\centerline{\psfig{file=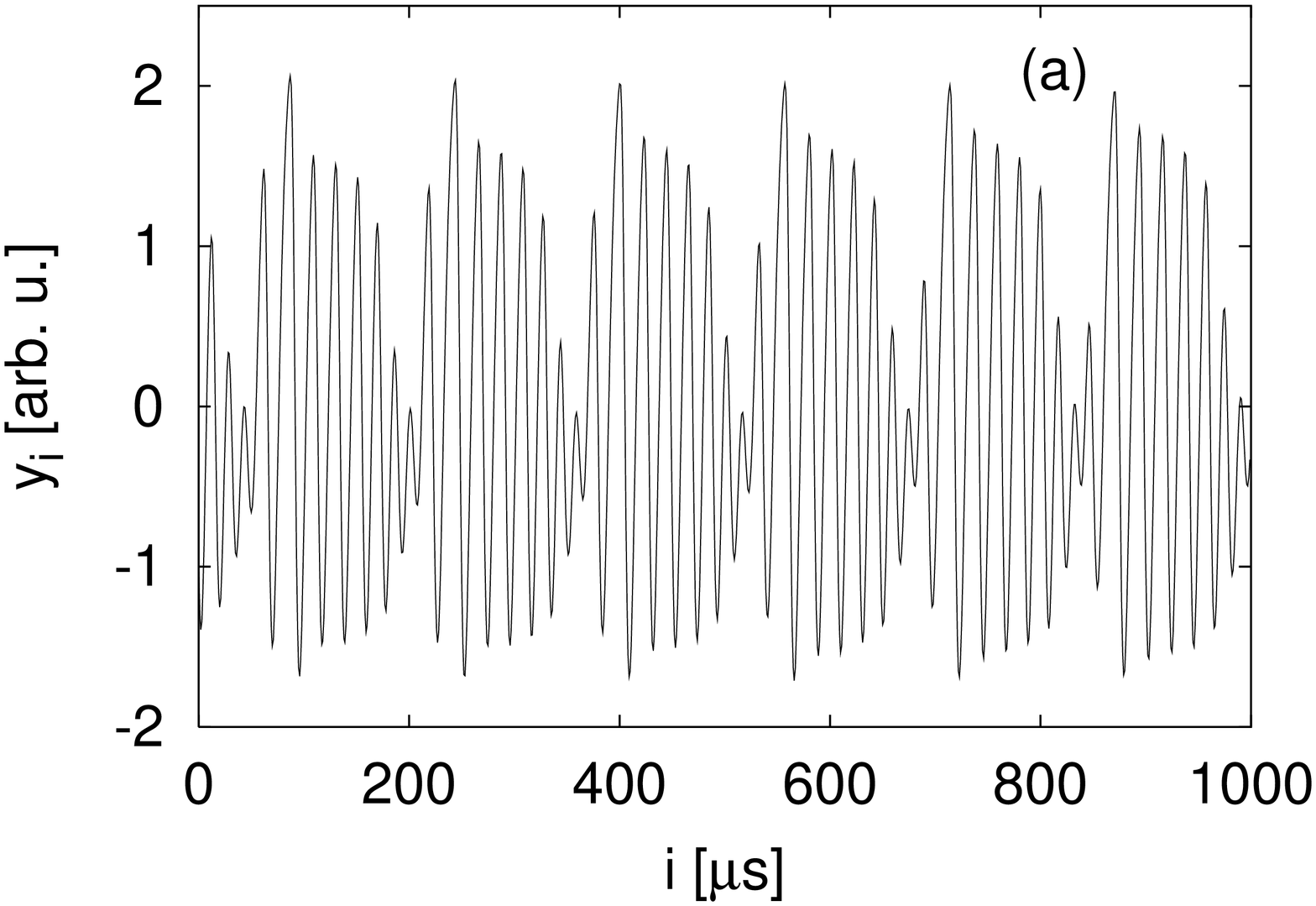,angle=270,width=7cm}
\psfig{file=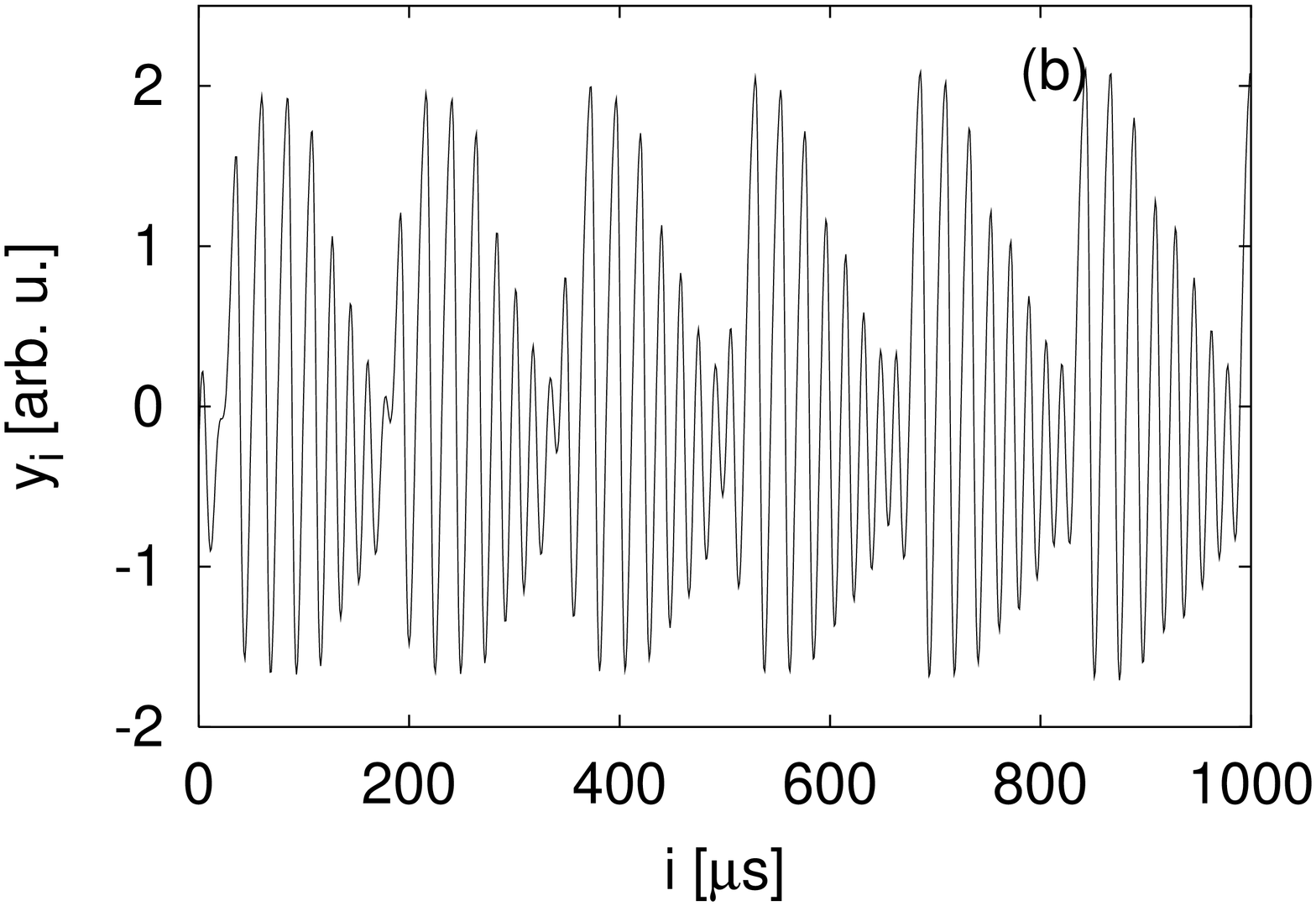,angle=270,width=7cm}}
\caption{First 1000 points of a trajectory with delay $\tau_b=150$
produced by using the data from the $\tau_a=400$ data set (panel a)
and the original data from the $\tau_0=150$ data set (panel b).}
\label{r400-150-ts.fig}
\end{figure}

\begin{figure}[ht]
\centerline{\psfig{file=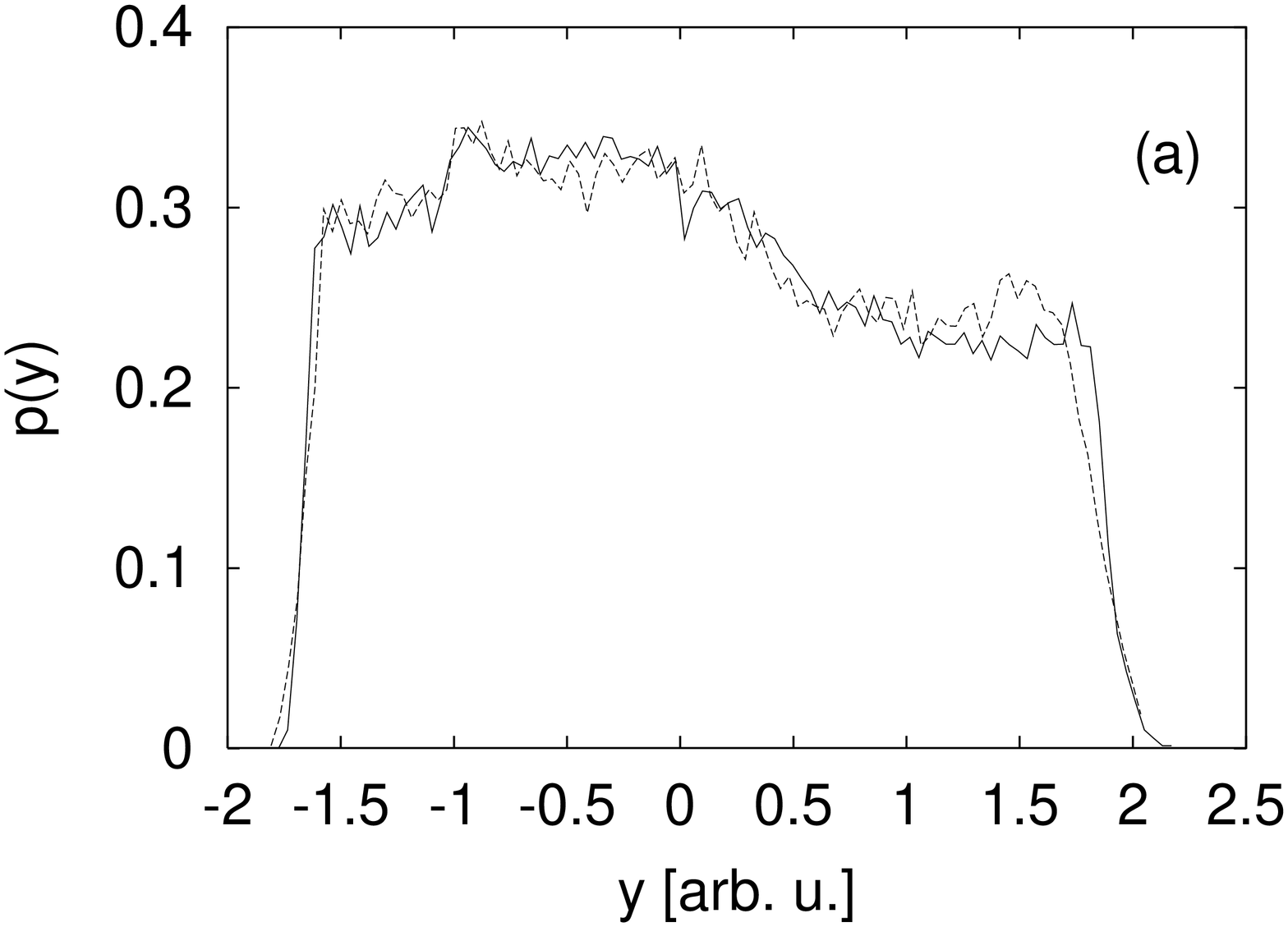,angle=270,width=7cm}
\psfig{file=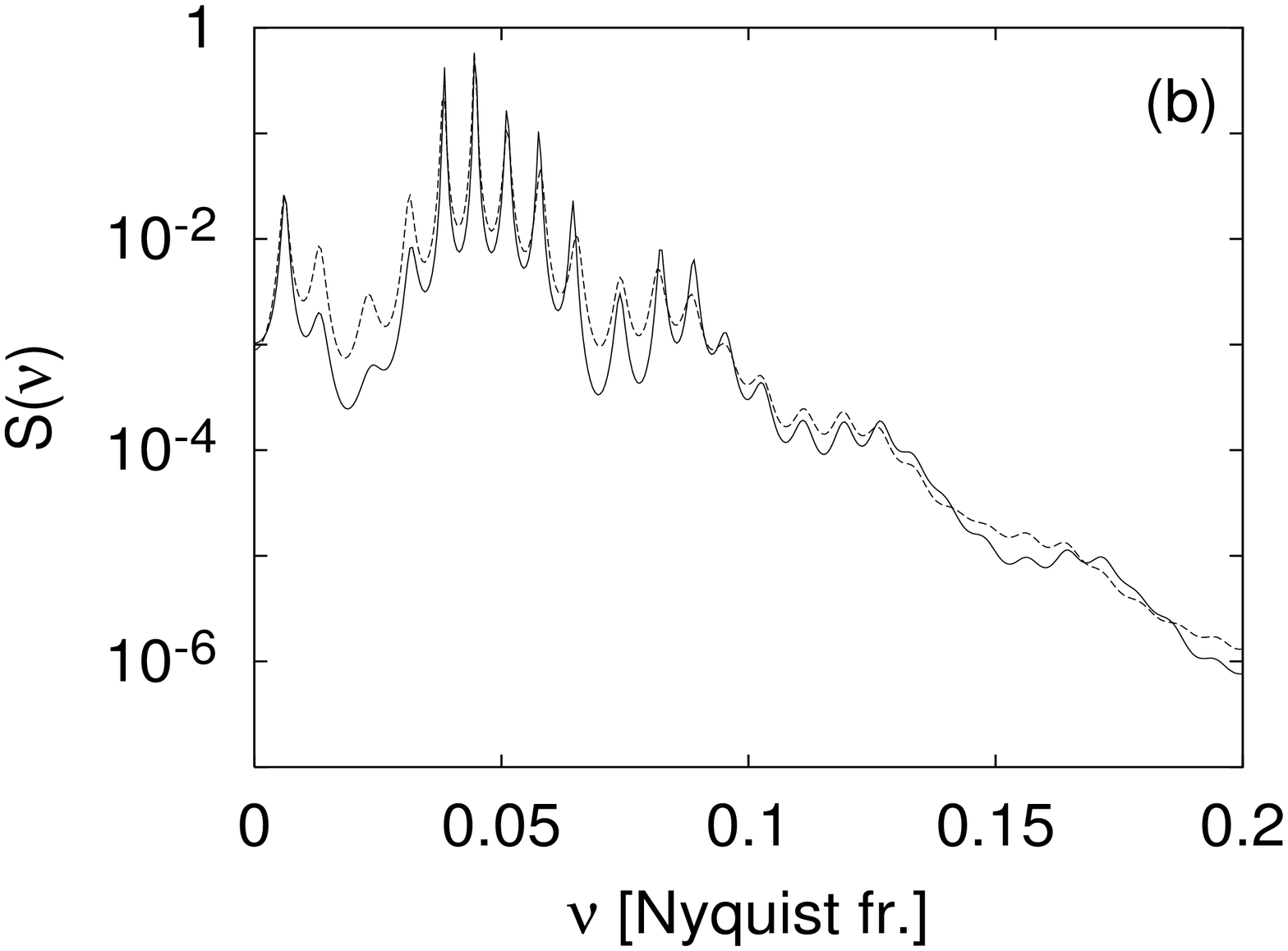,angle=270,width=7cm}}
\caption{Comparison of the invariant measure (more precisely the
scalar distribution) (panel a) and the power spectrum (panel b)
of the original data with $\tau_b=150$ (dashed line) and the data
produced by iterating the $\tau_a=400$ model with $\tau_b=150$ (solid
line).} 
\label{r400-150-his.fig}
\end{figure}

Figure~\ref{r400-150-ts.fig} shows 1000 points of the trajectory
produced by the above described procedure. The main visible feature is
a long periodicity approximately equal to $150 \mu s$ , which matches
the delay time $\tau_b$. Much more we cannot learn from this
figure. In order to perform a more quantitative comparison we have
looked at statistical properties of the data set. The left panel of
fig.~\ref{r400-150-his.fig} shows the scalar distribution of the data. 

The two lines shown in the plot are the distributions estimated from the
computer generated (solid line) and from the original (dashed line) 
trajectory with $\tau_b$. The agreement is certainly comparable with
that one guaranteed by the model constructed directly from the experimental
data for $\tau_b$ (including the absence of a peak around $y=-0.5$).
A similar conclusion can be drawn by looking at the power spectra in the
right panel of the figures. In particular, the various peaks that are
absent in the original dynamics at $\tau_a$ are located in the correct
positions. The most significant deviations are found in the low-frequency
region that is anyhow the most critical one as it can be observed in 
Fig.\ref{fig.r150-r400-spec}.

\begin{figure}[ht]
\centerline{\psfig{file=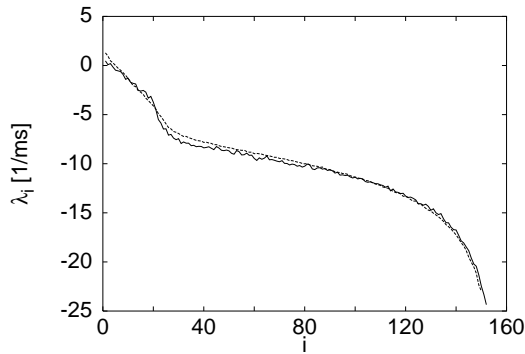,angle=270,width=7cm}}
\caption{Similar to fig.~\protect{\ref{r400-150-his.fig}}, but for the
Lyapunov spectra of both data sets.}
\label{r400-150-lyap.fig}
\end{figure}

As a last check we have computed the Lyapunov spectrum which
agrees all the way down to the smallest exponents 
(see Fig.~\ref{r400-150-lyap.fig}). Such a beautiful agreement must,
however, be interpreted as a proof of the stability of the method
(as we are comparing models obtained for two different delay times) rather 
than an indication of the correctness of the whole spectrum.

We also tried to apply the same procedure to generate a trajectory with
$\tau_0=50$. Unfortunately, all attempts have converged to a periodic orbit.
In our opinion, the reason is the retuning of the bias voltage B in
the experiment when switching from $\tau_0=50\mu s$ to $\tau_0=150\mu s$.
In the low-dimensional case, several periodic windows can be 
detected in chaotic regions. Thus, a small change in a control parameter
is likely to induce a transition either from chaotic to periodic
behaviour or vice versa.

\section{Conclusions}
\label{sec.conclusion}
We have analysed the behaviour of a CO$_2$ laser with a time-delay
feedback with a new embedding technique, designed to treat this class of
systems. For a short delay-time, where the dynamics is low-dimensional, 
we have been able to check our methodology by comparing its results with 
those of well established methods from nonlinear time series analysis. 
Already in that case, the approach proved to be superior, allowing to
construct a globally stable model. In the case of high-dimensional 
dynamics (longer delays) we have been able to model the system in a
state-space of dimension significantly smaller than that of the attractor. 
We could show that the minimal embedding dimension needed to reproduce all 
relevant features is $m_d=5$, indicating that the active degrees of freedom 
of the laser are bounded between 2 and 5. 

Furthermore, we have provided the first experimental evidence of the
scaling behaviour of the Lyapunov spectra with the delay time. Accordingly,
we have found that the dimension density is such that an additional active
degree of freedom is created when the delay time is increased by 16 $\mu s$.
Moreover, we have been able to confirm that the
the dynamical entropy does not increase with the delay time.

Finally, we want to stress that with the identification of a model in the
maximally chaotic regime (i.e. for sufficiently large delay) one could, in 
principle, study the whole bifurcation scenario of a delayed feedback system,
upon changing the delay time. One could, for example, study the transition from
standard chaos (one single positive Lyapunov exponent) to hyperchaos (more 
than one positive Lyapunov exponent) and compare with the experimental data. 
This interesting perspective is left to future investigations.

M.J.B. is supported by
a Marie-Curie-Fellowship of the EU with the contract number:
ERBFMBICT972305.; R. H. is partly supported by the EU with  contract
number: ERBFMRXCT96.0010 and wants to thank the colleagues at INOA
for their kind hospitality. Support from the EU project PSS 1043 is also
acknowledged.

\bibliographystyle{prsty}
\bibliography{co2b}

\end{document}